\documentclass{aa}  

\usepackage{graphicx}
\usepackage{txfonts}
\usepackage{hyperref}
\hypersetup{
    colorlinks=true,
    linkcolor=blue,
    filecolor=magenta,      
    urlcolor=blue,
    citecolor=blue,
    } 
\usepackage{lscape}
\usepackage{subfig}
\begin{document}

    \title{MeerKAT measurement of the radio emission from massive stars in the Galactic plane}
    \subtitle{I. Wolf-Rayet stars}

\titlerunning{WR with SMGPS}
\authorrunning{Tasseroul et al.}

   \author{M. Tasseroul
          \inst{1}
          \and
          M. De Becker\inst{1}
          \and
          P. Benaglia\inst{2}
          \and
          A. Tej \inst{3}
}
   \institute{Space Sciences, Technologies and Astrophysics Research (STAR) Institute, University of Liège, Quartier Agora, 19c,
Allée du 6 Août, B5c, 4000 Sart Tilman,
               Belgium \\
              \email{Mathilde.Tasseroul@uliege.be}
              \and
              Instituto Argentino de Radioastronomia (CONICET; CICPBA; UNLP), C.C. No 5, 1894, Villa Elisa, Argentina
              \and 
              Indian Institute of Space Science and Technology, Thiruvananthapuram 695 547, Kerala, India
}
   \date{Received 24 Avril 2026 / Accepted 9 July 2026}
 
  \abstract
   {Massive stars, including Wolf-Rayet (WR) stars, are predominantly found in binary systems. These systems are known to emit both thermal emission from stellar winds and, occasionally, non-thermal emission produced by relativistic electrons accelerated in the wind-wind interaction region.} 
   {We intend to provide the most complete census of radio emission from WR stars in the Galactic plane, using the SARAO Meerkat Galactic Plane Survey (SMGPS) complemented by the MeerKAT Galactic Center Survey (MGCS). Our main motivation is to identify hints of synchrotron radio emission indicative of particle-accelerating colliding-wind binaries (PACWBs).}
   {We compiled an input catalogue of 428 WR stars positionally covered by the SMGPS and the MGCS. Using the survey data, we measured the radio emission at 1.3 GHz for detected WR stars. We also measured the upper limits for objects located in sufficiently low radio background regions. For detected objects, we searched for a radio excess by comparing the measured fluxes to two different evaluators of the thermal emission from massive star winds.}
   {We detected 23 targets and determined the upper limits for 279 WR stars. Among the detected objects, 15 display a (significant or potential) radio excess that cannot be explained by unresolved circumstellar emission of any kind. After removing already known PACWBs, we report the identification of 12 potential new PACWB candidates.}
   {Our study has led to the compilation of the most extensive catalogue of Galactic WR radio (snapshot) emission to date, based on a homogeneous dataset covering the Galactic plane. The low detection rate indicates either a low occurrence rate of synchrotron emission or substantial attenuation by turnover processes that are clearly dominated by free-free absorption from the WR wind material. Our results open the door to dedicated follow-up observations aimed at ascertaining the nature of the identified radio excesses.}

   \keywords{radio continuum: stars -- 
                stars: massive --
                stars: Wolf-Rayet --
                Galaxy: stellar content}

   \maketitle
%

\section{Introduction}\label{Section:Intro}
The category of massive stars (M $\ge$ 8 $M_\odot$) encompasses a broad range of objects, including early B-type stars and O-type stars. Across their evolution, a fraction of them go through a Wolf-Rayet (WR) phase.
Massive stars are predominantly found in binaries or higher multiplicity systems \citep{Offner2023}. The most massive among them, in particular O-type and WR stars, possess powerful radiatively driven stellar winds, which are continuous outflows of material driven by radiation pressure. With respect to WR stars, which are the focus of this study, stellar winds are characterised by mass loss rates of the order of 10$^{-5}$\,M$_\odot$\,yr$^{-1}$ and terminal velocities of 1000 to 3000 km\,s$^{-1}$ \citep{CROWTHER2007}. This yields a very high wind kinetic power of the order of $10^{36}-10^{37}\, \mathrm{erg}\,\mathrm{s^{-1}}$, providing a significant energy reservoir for various physical processes. 

This paper is focussed more specifically on WR stars, which are distinguished by denser and more powerful winds, as well as the presence of strong, broad emission lines in their spectra. Their spectral classification is based on the dominant chemical elements observed in these emission lines, resulting in three main spectral types \citep{CROWTHER2007}: WN (nitrogen and helium lines), WC (carbon and helium lines), and WO (oxygen lines). 

Massive stars in binary systems exhibit two types of radio emission. The first is thermal Bremsstrahlung emission from their individual ionised stellar wind, at a temperature that is of the order of $10^4$\,K \citep[]{WrightBarlow,PanagiaFelli}. Systems composed of massive stars whose winds interact are referred to as colliding-wind binaries (CWB). The second emission process is non-thermal synchrotron emission, arising from relativistic electrons accelerated in the wind-wind interaction region of CWB, in the presence of stellar magnetic fields \citep{EichlerUsov1993}. The collision of these hypersonic winds produces high-Mach-number shocks, which accelerate electrons through the diffusive shock acceleration (DSA) mechanism \citep{Drury1983}. In particular, CWBs that exhibit non-thermal emission are classified as particle-accelerating colliding-wind binaries (PACWBs). They have been compiled in a dedicated catalogue that includes, to date, 61 objects \citep{CataloguePACWB}\footnote{The online catalogue is available at \url{https://www.astro.uliege.be/~debecker/pacwb/home.html}}, 25 of which are WR stars.

When the colliding-wind region is not spatially resolved from the stellar winds, PACWBs display a composite radio spectrum with contributions from both thermal and non-thermal components. Thermal emission follows a power-law spectrum with a positive spectral index: for smooth, isothermal, and isovelocity winds, the flux density scales as $S_{\nu} \propto \nu^{0.6}$ \citep{WrightBarlow}. However, for WR stars, observations show that the spectral index is steeper, with values of 0.7-0.8 instead, likely due to their denser and clumpier winds \citep{Nugis1998}. In contrast, synchrotron emission follows a rather flat or even negative power-law index, making it dominant at lower frequencies.
Several turnover processes could potentially affect the synchrotron spectrum at low frequencies, leading to attenuation or even complete suppression of the non-thermal signature in the radio spectrum. The main turnover process active in PACWBs is free-free absorption (FFA), which consists of the absorption of the synchrotron emission by thermal electrons present in the dense wind. Understanding this mechanism is crucial, as it can mask synchrotron signatures and thus hinder the identification of new PACWBs \citep[]{DeBecker2019,Arora2021,Blanco2024}. In addition to FFA, the Razin-Tsytovitch effect might, in principle, play a role below 1\,GHz \citep{Pacholczyk1970,Pittard2005}, while synchrotron self-absorption would require the synchrotron emission region to be more compact than in the case of PACWBs for it to be significant. Any turnover occurring close to (or above) 1\,GHz is expected to be dominated by FFA, given the high density of the wind material within and around the synchrotron emission region. 

To date, most radio observations of massive stars have focussed on specific targets or clusters, leading to the identification of PACWBs based on deviations from expectations for pure thermal emission \citep[e.g.][]{SetiaGunawan2003,Benaglia4stars,AndrewsWesterlund,Benaglia2020Cygnus,CanoGonzArches,CanoGonz}. However, apart from these dedicated studies, a general and broader sampling of the radio emission of massive Galactic stars is still lacking.
The SARAO MeerKAT Galactic Plane Survey (SMGPS) was released in June 2024, providing the largest and most sensitive radio survey at 1.3 GHz with a high angular resolution of 8 arcseconds \citep{SMGPS}. It is also complemented by the MeerKAT Galactic Center Survey (MGCS) which covers the Galactic centre region at the same frequency but with an angular resolution of 4 arcseconds \citep{GalacticCenter}. One of our objectives is to build a catalogue of radio fluxes for massive stars in the Galactic plane, based on both MeerKAT surveys, which will enable a broader view of the radio emission properties of these stars. Our second aim is to understand the nature of their radio emission, distinguishing between purely thermal and potential non-thermal sources, and therefore identifying PACWB candidates. This first paper is focussed on WR stars, while a follow-up study will address the case of O-type stars.

The paper is organised as follows. The compilation and content of the input WR star catalogue are provided in Sect.\,\ref{Section:WR catalogue}. The radio measurements of the WR star in the input catalogue are described in Sect.\,\ref{Sect:measurements}. In Sect.\,\ref{Section: Results}, we present the main results in terms of detection statistics. Section \ref{Section: Discussion} addresses the question of the nature of the detected radio emission, leading to potential PACWB candidates. We summarise our conclusions in Sect.\,\ref{Section: Conclusions}.

\section{The Wolf-Rayet catalogue}\label{Section:WR catalogue}
An overview of the WR catalogue compliant with the sky coverage of SMGPS and MGCS is provided in Table \ref{Tab:CompleteCatalogue}. This input catalogue was compiled using several previously published resources, including the Galactic Wolf Rayet Catalogue from \citet{CrowtherCatalogue}\footnote{\url{https://pacrowther.staff.shef.ac.uk/WRcat/index.php}}(v1.31), which includes the members of the recent catalogue of potential new WR stars identified in {\it Gaia} DR3 data by \citet{Mulato}, and the SIMBAD database \citep{SIMBAD} queried via the Table Access Protocol (TAP) to extract WR-type stars only. 
To match these sources with the SMGPS and MGCS data, the stars were filtered by Galactic longitude and latitude, selecting only those within the MeerKAT survey coverage: $251\degr \le l \le 358\degr$ and $2\degr \le l \le 61\degr$ at $|b| \le  1.5\degr$ for the SMGPS and between $l = 358\degr$ and $l = 2\degr$ for the MGCS. All the WR stars in the aforementioned catalogues located in these ranges were included.
In total, the input WR catalogue includes 428 stars.
 
The first column of the catalogue lists the star identifier. The second column provides the spectral type. If the star appears in either the Galactic WR catalogue or the Gaia-based catalogue by \citet{Mulato}, the spectral type was taken from these dedicated catalogues. Columns three to six contain the coordinates of the stars. The right ascension and declination were retrieved from SIMBAD and transformed into Galactic longitude and latitude using the SkyCoord Python package included in \citet{astropy}. The stars are listed by increasing Galactic longitude. The seventh column gives the distance to each star, expressed in parsecs. Distances were compiled from multiple sources in the following order of priority. First, we used the Gaia Data Release 3 (GDR3) distances revised by \citet{Crowtherdistance} (Gaia2023 in column 8). Second, we adopted the regular Gaia DR3 distances reported by \citet{GDR3distance}  (Gaia2021 in column 8). It is important to note that distances from GDR3 may not be accurate because parallax-based distances are generally less reliable for WR stars. Their high luminosity and strong stellar winds can affect the astrometric solution and degrade the accuracy of the measurements \citep{DingpbdistanceGDR3}. This is why a fall-back solution was to consider the distances reported in the Galactic Wolf–Rayet Catalogue \citep{CrowtherCatalogue} for the remainder (about 130 targets), due to a lack of a Gaia counterpart with a reasonable relative error on the distance (WRCat in the 8th column). In some cases, the lack of an existing WRCat distance meant that we needed to quote the Gaia DR3 value despite a large relative error on the distance. These objects are marked with a  `*' next to their distance value. We also note that for one specific target, the only distance is that proposed in its original discovery paper \citep{SFZ2012}. The ninth column indicates which source was used to compile the full input catalogue: WRCat \citep{CrowtherCatalogue} or SIMBAD \citep{SIMBAD}.
\section{Flux density measurements}\label{Sect:measurements}

The input catalogue described in Sect.\,\ref{Section:WR catalogue} was used to search for radio counterparts in SMGPS and MGCS data. For each WR position, the primary beam-corrected images were downloaded and scrutinised to identify detected sources and make a census of non-detections, with careful consideration given to the existence of diffuse or unresolved sources at each specific position. 

\begin{table*}
\caption{Catalogue of selected Wolf-Rayet stars sorted by increasing Galactic longitude (sample).}
\label{Tab:CompleteCatalogue}
\centering
\scriptsize
\begin{tabular}{lccccccccccc}
\hline\hline
Name & Sp. type & RA & Dec & $l$ & $b$ & Distance & Dist. source & Main source &$S_{1.3}$ &
$\sigma_{S_{1.3}}$ & Flag \\ 
 & & ($^\circ$) & ($^\circ$) & ($^\circ$) & ($^\circ$) & (pc) & & & (mJy) & (mJy) & \\
\hline
...&...&...&...&...&...&...&...&...&...&...&... \\
WR 102-19 & WN5 & 268.83420 & -24.12732 & 5.24270 & 0.59906 & 4499 & Gaia2021 & WRCat & $\leq 0.59$ & / & UL\\
WR 102-20 & WC9 & 269.76186 & -24.34740 &  5.47647 & -0.24315 & 1487 & Pre-Gaia & WRCat & / & / & HB\\
WR 102-22 & WC7; WC7d & 269.75975 & -24.28334 & 5.53104 & -0.20954 & 3729 & Gaia2021 & WRCat & $\leq 0.18$ & / & UL\\
WR 104 & WC9d+B0.5V (+VB) & 270.51719 & -23.62838 & 6.44319 & -0.48541 & 670 & Gaia2021 & WRCat & $\leq 0.53$ & / & UL\\
WR 105 & WN9h & 270.59774 & -23.57708 & 6.52425 & -0.52427 & 3500 & Gaia2023 & WRCat & 0.43 & 0.15 & S \\
WR 102l & WN8o & 270.14304 & -22.79443 & 6.99747 & 0.22628 & 3184 & Gaia2021 & WRCat & $\leq 1.47$ & / & UL\\
WR 108 & WN9ha & 271.35723 & -23.00565 & 7.36476 & -0.85110 & 3100 & Gaia2023 & WRCat & $\leq 0.28$ & / & UL \\
WR 102-21 & WN6 & 269.95094 & -22.24783 & 7.38369 & 0.65184 & 6789 & Gaia2021 & WRCat & / & / & HB\\
...&...&...&...&...&...&...&...&...&...&...&... \\
\hline
\end{tabular}
\tablefoot{Columns: Name of the star, spectral type from the reference catalogue, right ascension (RA), declination (Dec), Galactic longitude ($l$), Galactic latitude ($b$), distance (pc), source for the distance (Gaia2023, Gaia2021, Pre-Gaia, or SFZ2012), main source (WRCat or SIMBAD),} flux density at 1.3\,GHz in mJy and associated error in mJy and a flag where HB stands for high background, UL for upper limit, S for single source associated with a unique target and U for unresolved targets associated with a unique source.
\end{table*}

\begin{figure}[h]
\centering
\includegraphics[width=1.0\linewidth]{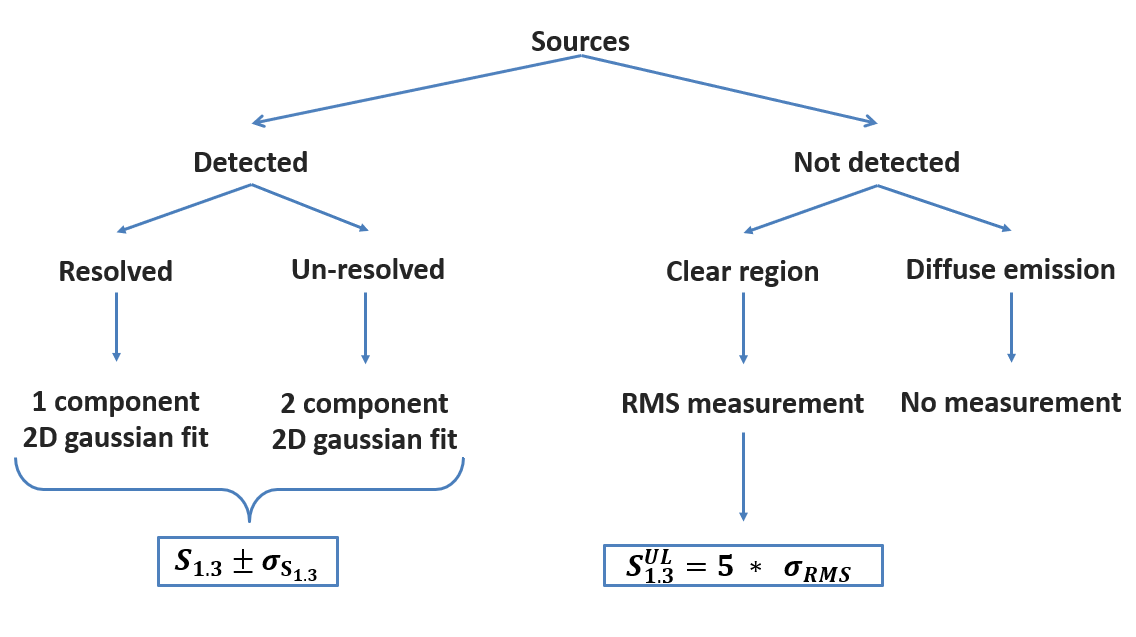}
\caption{Decision tree adopted for the radio measurements.}
\label{Fig:decision_tree}
\end{figure}

The results of our radio measurement procedure are listed in the last two columns of Table\,\ref{Tab:CompleteCatalogue}. Different cases were encountered during the analysis of the images. The methods adopted for each case are presented in Fig\,.\ref{Fig:decision_tree} and described in detail below. For detected point sources without significant contamination by nearby emission, the flux density at 1.3 GHz and the associated error were obtained using the \texttt{Image fitting} task available in \texttt{CARTA} (Cube Analysis and Rendering Tool for Astronomy)\footnote{\url{https://cartavis.org/}}. This tool fits a 2D Gaussian component to the source based on an initial estimate of its centre, amplitude, and full-width-half-maximum (FWHM) with a given solver. In this work, we used the Cholesky solver\footnote{\url{https://carta.readthedocs.io/en/latest/image_fitting.html}}. Given the non-zero background in the vicinity of each source, we took into account a fixed background component in the fit. The adopted background level was taken as the average of several regions close to the source which we assumed to the representative of the actual background at the source position. For the specific case of a point source of interest partly blended with another one, a two-component 2D Gaussian fit was performed. The output of this procedure is a measure of the flux density with standard deviation for each detected point source among members of the input WR catalogue.

For non-detected targets, provided that the location of interest was not dominated by bright diffuse emission, we measured the root-mean-square (rms) noise ($\sigma_\mathrm{RMS}$) in a circular region with a radius of 4 arcseconds (the typical size of the synthesised beam) close to the position of the target coordinates. We adopted conservative upper limits on the flux density as $S_\nu^\mathrm{UL} = 5\,\sigma_\mathrm{rms}$. However, for several targets, the estimate of a relevant upper limit was more complex due to large scale and bright diffuse emission contaminating the source position. In many cases, the background emission is at the 1--10 mJy level, which is too bright to retrieve any relevant information from any embedded faint source. In these circumstances, we refrained from proposing speculative upper limits on the emission from these specific targets.

\section{Results}\label{Section: Results}
Among the input WR catalogue introduced in Sect.\,\ref{Section:WR catalogue}, a total of 23 targets display clear point-like radio emission resolved from any neighbouring diffuse emission (corresponding to 5.4\,\% of the input catalogue). Accounting to the fact that 124 targets are embedded within larger structures with high radio background, the fraction of detected objects among sufficiently low background regions is 7.6\,\%. We note that three pairs of targets (WR\,84-6 and WR\,84-7, WR\,46-3 and WR\,43-4, and WR\,76-6 and WR\,76-7) are unresolved at the angular resolution of the survey, making it difficult to associate the observed emission with either star. They were therefore considered a single entry in the remainder of the paper (WR\,84-6/7, WR\,46-3/4 and WR\,76-7/8, respectively). In the input WR catalogue, these objects are denoted by ‘U’ in the flag column. We note that WR\,76-6 is also located close to the WR\,76-7/8 pair but with an offset that is large enough to be resolved. Finally, for a total of 278 WR stars (65.5\,\% of the full input catalogue and 92.4\,\% of targets located in low background emission regions), upper limits have been determined following the methodology described in Sect.\,\ref{Sect:measurements}. The radio images of the 23 detected sources are presented in Fig.\,\ref{fig:radio_images}. 

The majority of WR stars located in the MGCS are embedded in the region dominated by the strong diffuse radio emission from the surroundings of Sagittarius\,A* \citep[]{GalacticCenter}. As explained in Sect.\,\ref{Sect:measurements}, this stands in the way of a valid extraction of the flux density. Consequently, apart from a few upper limits (WR  98a, [SFZ2012] 1269-166L, [SFZ2012] 1275-184L
), no relevant measurement could be taken for this subset of targets.

\section{Discussion}\label{Section: Discussion}
\subsection{Origin of the low detection fraction}\label{lowdetection}
\begin{figure}[h]
\centering
\includegraphics[width=1.0\linewidth]{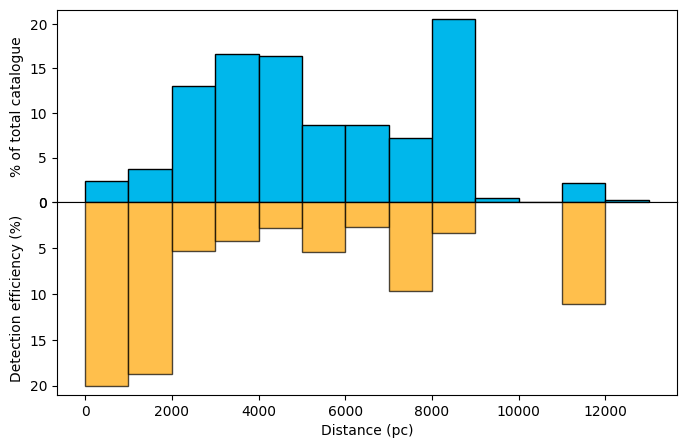}
\caption{Distance distribution of the members of the input WR catalogue. Top: Histogram showing the distribution of the population of WR stars in the input catalogue as a function of distance. Bottom: Histogram showing the detection efficiency as a function of distance, defined as the ratio between the number of detected stars and the total number of catalogue stars in each distance bin.}\label{Image: distance distribution}
\end{figure}

According to the statistics presented in Sect.\,\ref{Section: Results}, more than 90\,\% of WR stars included in the catalogue have not been detected. We must consider that these non-detections might arise from sources that are not intrinsically bright enough to lead to a valid detection due to the geometrical dilution of the flux density that scales with the inverse of the square of the distance. The distribution of WR distances, displayed in Fig.\,\ref{Image: distance distribution}, shows that most of the objects are located between 2 and 5 kpc, with a peak near 8 kpc corresponding to the MGCS population. 

As stated in Sect.\,\ref{Section:Intro}, at arcsecond resolution, the broadband radio spectrum of massive stars is either purely thermal in the absence of measurable synchrotron radiation or composite. In the former case, the measured radio emission is purely optically thick thermal free-free emission that is directly dependent on the wind properties of the star, as developed in Sect.\,\ref{Section:WrightBarlow}. The likelihood of detecting such sources increases for shorter distances and for winds strong enough to be bright in thermal radio emission. Depending on the local background, potentially faint sources may still be undetected. The occurrence of pure thermal emission is expected from a single WR star or from a binary from which no synchrotron emission has been measured. This can be due either to a lack of particle acceleration or to a strong attenuation of the synchrotron emission by FFA.

The FFA coefficient is proportional to the square of the density of the wind material \citep{WrightBarlow,PanagiaFelli} and WR stars typically exhibit stellar winds that are much denser than their O-type progenitors. For either FFA from foreground dense wind material or internal FFA, due to a thermal plasma that coincides at least partially with the synchrotron emission region \citep{Tasseroul2025}, the synchrotron emission component (if any) can be significantly attenuated and even suppressed in extreme cases, especially at low frequencies \citep{Benaglia2019,DeBecker2019,Saha2023,Blanco2024}. As a consequence, WR winds display two competing properties: their high kinetic power favours synchrotron radio emission, whereas their high density inhibits its detection by enhancing absorption. 

\subsection{Known WR-type PACWB}\label{Sect: knownPACWB}
The PACWBs in list A of \citet{CataloguePACWB}\footnote{\url{https://www.astro.uliege.be/~debecker/pacwb/listA.html}} were also checked in the SMGPS images for detections. Ten of them fall within the longitude and latitude range of the SMGPS: WR\,104, WR\,105, WR\,112, WR\,125, WR\,21a, WR\,39, WR\,70-16, WR\,77o, WR\,89, and WR\,98. Among these, we see that WR77o is blended within a large radio structure, certainly due to the combined emission from many stellar sources in the Westerlund\,1 cluster. The radio emission associated with WR\,104, WR\,125, WR\,21a, and WR\,89 displays upper limits of 0.53, 0.74, 11.67, and 5.06 mJy, respectively. The remaining stars, WR\,105, WR\,112, WR\,39, Apep, and WR\,98, were detected with flux densities of 0.82, 2.50, 1.86, 167.62, and 0.2 mJy, with respective uncertainties of 0.27, 0.05, 0.06, 0.22, and 0.02 mJy. One particular case is worthy of note in this context: we see that the coordinates of WR\,39 are offset by about 3 arcseconds from the centre of the detected point source. While the emission is significant, we cannot fully reject the scenario of radio emission from a nearby or line-of-sight object.

These detected PACWBs have been classified as non-thermal emitters based on their radio emission alone \citep[]{VanderHucht2001,Marchenko2002,AndrewsWesterlund,Apep}, except for WR\,70-16 (Apep), which was also identified with hard X-ray non-thermal emission \citep{ApepXrays}. The prevalence of synchrotron radio emission in the identification of particle accelerators reinforces the relevance of using radio surveys, such as the SMGPS, to find new PACWBs in a significantly large sample. In addition, \citet{CataloguePACWB} proposed a supplementary list of objects with uncertain status, which could turn out to be PACWBs following a confirmation by additional observations. WR\,79 (HD\,152270) is one of these objects. However, it has not been detected in SMGPS images.

Finally, we note that some known PACWBs are members of the Arches and Quintuplet clusters \citep{CanoGonzArches,CanoGonz} and they are therefore covered by the MGCS. Due to the very high background level in that region at the frequency of the survey, none of them could be properly measured.

\subsection{The nature of the detected radio emission}
Knowing only the flux densities of the stars is not sufficient to determine the nature of their emission (thermal or non-thermal). The approach to identifying it consists of comparing the measured flux densities with estimates of the thermal emission. If the measured value significantly exceeds the theoretical prediction, a radio excess can be identified, which might indicate non-thermal radiation. In this work, two methods were employed to estimate the thermal emission of these stars, which are presented below. 

\subsubsection{Theoretical prediction method}\label{Section:WrightBarlow}
The theoretical prediction of thermal emission for a massive star wind can be obtained using the formalism of \citet{WrightBarlow}, which provides the flux density as a function of frequency, mass-loss rate, terminal velocity, and distance. However, as mentioned previously, the authors considered several assumptions, including homogeneous winds with constant velocity and smooth winds. This last hypothesis is not representative of WR stars, whose winds present clumps instead. 
To account for this effect, we used a modified formula that includes a wind volume filling factor, $f_w$, which is equivalent to the inverse of the clumping factor in the framework of the void interclump hypothesis \citep{Puls2006}. This approach is in agreement with previous studies in which a clumping correction was provided to the smooth wind theory \citep[][and references therein]{Saha2023,Blanco2024} via
\begin{equation}\label{Eq1}
    S_{1.3}^\mathrm{Th} = 23.2 \bigg(\frac{\dot M}{v_\infty \sqrt{f_w}\, \mu}\bigg)^{4/3} \, \bigg(\frac{\gamma\,g_{ff}\, Z^2 \, \nu}{D^3}\bigg)^{2/3} .
    \end{equation}

Equation \ref{Eq1} expresses the flux density $S_{1.3}^{\mathrm{Th}}$ in Jy, which is specifically computed at a frequency of 1.3\,GHz), the mass loss rate $\dot M$ in M$_\odot\,\rm{yr^{-1}}$, the terminal velocity $v_\infty$ in km\,s$^{-1}$, the distance $D$ in kpc, and the volume filling factor, $f_w$, set to 0.25. Regarding the plasma parameters, the rms ionic charge, $Z$, and the mean number of electrons per ion in the wind were set to 1 for WN stars; whereas we used 1.5 for WC stars for both. These values are in fair agreement with numbers adopted by previous studies \citep{ClausII,Montes2009}. For the mean molecular weight, we selected values inspired by the results arising from the PoWR atmosphere code \citep{PoWR2002,PoWR2003,PoWR2015}. One must, however, caution that the PoWR convention consists of using the total mean molecular weight (including both ions and electrons, $\mu_{\rm{tot}}$), while the \citet{WrightBarlow} formalism assumes only ions. Both conventions are related such that $\mu = (1 + \gamma)\,\mu_{\rm{tot}}$. For WC stars, the PoWR profiles suggest $\mu_{\rm{tot}} \sim 2.2$ in the outer wind, leading to $\mu \sim 5.5$. For WN winds, following the same approach, we adopted $\mu_{\rm{tot}}$ = 1 and 2 for late-type and early-type WN stars, respectively, leading to $\mu$ = 2 and 4. The free-free Gaunt factor, $g_{ff}$, is expressed following \citet{gauntfactor_Letiherer} via
\begin{equation}
    g_{ff} = 9.77 \, \bigg(1+0.13\, log\frac{T_e^{3/2}}{Z\nu}\bigg).
\end{equation}
\noindent Here, $T_e$ is the electron temperature of the wind in its outer part, where the free-free emission originates. The dependence of the thermal radio emission on the electron temperature is rather weak, as it only affects the radio flux density through the Gaunt factor, where its influence is damped by the logarithm function. According to the temperature profiles obtained with the PoWR code (see references above), in the outer wind, $T_e$ is expected to reach values as low as 8000--15000\,K, depending on the WR type. This is in fair agreement, for instance, with the 9400 K value recommended by \citet{Morris2000} for late-type WN winds, or 8000 to 9000\,K for WC stars \citep{Hillier1989,Dessart2000}. Given the weak dependence on $T_e$, adopting a value of 10$^4$ for all WR types is valid and somewhat better than opting for the $T_e \approx 0.3\,\times\,T_{eff}$ prescription that is frequently used for O-type stars \citep{Drew}. To give an idea of the impact of the decision on the electron temperature, the relative difference in $g_{ff}$ between 8000 and 14000 is about 8\,\%. Given the $g_{ff}^{2/3}$ dependence of the flux density (see Eq.\,\ref{Eq1}), this results in a relative difference in $S_{1.3}^\mathrm{Th}$ of about 5\,\%, which is well below the conservative relative error on the predicted flux density discussed below.

For detected stars discussed in detail in Sect.\,\ref{Section:Excess}, this formula was applied to derive model sources with thermal emission from their winds. The stellar parameters were taken from \citet{HamannWN} for the WN stars and from \citet{SandersWCWO} for the WC and WO stars. If a star was included in one of these lists, we adopted the published parameters directly. Otherwise, we used the average mass-loss rate and terminal velocity of stars with the same spectral subtype. In \citet{SandersWCWO}, the mean per subtype is already provided in their Table\,5, while for \citet{HamannWN} we calculated them ourselves for the subtypes involved. All these values are reported in Table\,\ref{tab:stellar_param_WR}.

\begin{table}
\caption{Stellar parameters of the WN and WC stars}
\label{tab:stellar_param_WR}
\centering
\begin{tabular}{lccc}
\hline\hline
Type &
$T_{eff}$ (K) &
$v_\infty$ (km\,s$^{-1}$) &
$\log \dot{M}$ (M$_\odot$\,yr$^{-1}$) \\
\hline
WN6 & 57242 & 1643 & -4.525 \\
WN7 & 53277 & 1368 & -4.523 \\
WN8 & 43108 & 997  & -4.267 \\
WN9 & 37650 & 985  & -4.650 \\
\hline
WC5 & 83000 & 2780 & -4.39 \\
WC6 & 78000 & 2270 & -4.47 \\
WC7 & 71000 & 2010 & -4.57 \\
WC8 & 60000 & 1810 & -4.53 \\
WC9 & 44000 & 1390 & -4.66 \\
\hline
\end{tabular}
\tablefoot{WN values from \citet{HamannWN}. WC values from \citet{SandersWCWO}.}
\end{table}

We estimated the uncertainty in the predicted thermal flux density using Eq.\,\ref{Eq1}, where we take into account the uncertainty in $\dot M$, $v_\infty$, $f_\mathrm{cl}$, $\mu$, and $D$, via
 
\begin{equation}
\begin{split}
\left(\frac{\sigma_{S_{1.3}^\mathrm{Th}}}{S_{1.3}^\mathrm{Th}}\right)^2
&=
\frac{16}{9}\left[
\left(\frac{\sigma_{\dot M}}{\dot M}\right)^2
+
\left(\frac{\sigma_{\mu}}{\mu}\right)^2
+
\left(\frac{\sigma_{v_\infty}}{v_\infty}\right)^2
+
\left(\frac{\sigma_{Z}}{Z}\right)^2
\right] \\
&\quad
+
\frac{4}{9}\left[\left(\frac{\sigma_{f_{w}}}{f_{w}}\right)^2 + \left(\frac{\sigma_{\gamma}}{\gamma}\right)^2\right]
+
4\left(\frac{\sigma_D}{D}\right)^2.
\end{split}
\end{equation}

For most of these parameters, it was not possible to evaluate physically relevant specific uncertainties for all targets. We therefore assumed typical relative errors for these parameters: $\sigma_{\dot M}/{\dot M} \approx 0.50$, $\sigma_{v_\infty}/v_\infty\approx 0.10$, $\sigma_{f_{w}}/f_{w}\approx 0.33$, $\sigma_{\mu}/\mu\approx 0.33$, $\sigma_{\gamma}/\gamma\approx 0.2$, $\sigma_{Z}/Z\approx 0.2$, and $\sigma_D/D\approx 0.20$. It is reasonable to consider that the uncertainty on $S_{1.3}^\mathrm{Th}$ is dominated by that on the mass loss rate, given the spread of ${\dot M}$ values reported for WR stars of a given sub-type, while the spread of terminal velocities is much lower \citep{HamannWN,SandersWCWO}. The assumed 33\,\% uncertainty on the volume filling factor converts into a clumping factor (1/f$_{w}$) ranging between 3 and 6, compliant with typical outer wind values predicted by \citet{RO2002}. For $D$, a 20\,\% uncertainty is very conservative in some cases and more typical of Gaia results for most of them. However, it is still not a dominant contribution to error propagation. The range of values found in the literature for $\mu$ \citep{ClausII,Montes2009}, even for WRs of similar subtypes, is well reflected by the adopted relative error of 33\,\%, and the same is true for $Z$ and $\gamma$, with a quite conservative value of 20\,\%. 

With the relative errors assumed above, we obtained a typical and conservative relative error on the predicted thermal flux density $\sigma_{S_{1.3}^\mathrm{Th}}/S_{1.3}^\mathrm{Th} \approx 0.98$. The resulting thermal flux density predictions ($S_{1.3}^\mathrm{Th}$) and their typical error will be used in Sect.\,\ref{Section:Excess}. It should be noted that WR\,46-3 and WR\,46-4 are considered as a single object in our detections. However, they have different spectral types. We therefore conservatively adopted the stellar parameters, leading to the highest theoretical estimate.

\subsubsection{Empirical extrapolation method}\label{Section: Estimation_avec_ref}
The model developed by \citet{WrightBarlow} predicts a spectral index of 0.6; however, as mentioned above, WR stars often exhibit steeper spectral indices \citep{Nugis1998}. In addition, the simple use of the \citet{WrightBarlow} model sometimes leads to predictions that are significantly lower than the actual measurements \citep{Saha2023,Blanco2024}. In the absence of any quantitative model valid for WR winds, we adopted an alternative approach, namely, an empirical method based on a chosen reference star of the same spectral subtype. Specifically, we compared the flux density of a reference star, rescaled to the actual distance of the target, with the measured flux density of the target that shares the same spectral type. We note that these reference stars were selected from a clear diagnostic of pure thermal radio emission based on flux densities measured at frequencies of several GHz. The fact that some of these objects are likely binaries is not an issue. Although the identification of synchrotron radiation is an indicator of binarity, binarity is not enough to warrant the measurement of synchrotron radiation. As clarified in Sect.\,\ref{lowdetection}, the non-thermal emission region (if any) may be too deeply buried in the WR wind to allow for a measurable amount of synchrotron radiation to escape if the orbital period is not long enough. As a result, our selection of so-called thermal reference stars is not corrupted by non-thermal contamination.

The reference values for each subtype were taken from \citet[]{Nugis1998,Montes2009,ClausI,ClausII}, in which we selected the flux densities at 4.8 GHz and 8.6 GHz for the stars considered pure thermal emitters at those frequencies. In a second step, we computed the spectral luminosity at 4.8\,GHz or 8.6\,GHz for each reference source as 
\begin{equation}
    L_\nu = 4 \pi D^2 S_\nu \,,
\end{equation}
where $D$ is the distance of the star. We selected reference sources that display a thermal index at the frequencies mentioned above to prevent our reference sources of thermal emission from being contaminated by synchrotron emission.

\begin{table*}
\caption{Flux densities and spectral luminosities at 4.8 and 8.6\,GHz for the pure thermal-emitting reference stars.}
\label{tab:thermal_references}
\centering
\scriptsize
\begin{tabular}{llcccccccccccc}
\hline\hline
Name &
Sp type &
$S_{4.8}$ &
$\sigma_{S_{4.8}}$ &
$S_{8.6}$ &
$\sigma_{S_{8.6}}$ &
$D_{ref}$ &
$\sigma_{D}$ &
$L_{4.8/8.6}$ &
 $\sigma_{L_{4.8/8.6}}$ &
$L_{4.8/8.6}$ &
$\sigma_{L_{4.8/8.6}}$ &
$\nu L_{\nu}$ &
Ref. \\
 &
 &
(mJy) &
(mJy) &
(mJy) &
(mJy)&
(pc) &
(pc)&
(erg\,s$^{-1}$\,Hz$^{-1}$)&
(erg\,s$^{-1}$\,Hz$^{-1}$)&
(Jy\,pc$^{2}$)&
(Jy\,pc$^{2}$)&
(erg\,s$^{-1}$)&
 \\
\hline
WR144 & WC4 & 0.67 & 0.25 & / & / & 1700 & 50 & $2.32\times10^{18}$ & $8.75\times10^{17}$ & $2.43\times10^{4}$ & $9.19\times10^{3}$ & $1.11\times10^{28}$ & NUGIS$^{a}$ \\
WR111 & WC5 & 0.33 & 0.10 & / & / & 1300 & 50 & $6.67\times10^{17}$ & $2.09\times10^{17}$ & $7.01\times10^{3}$ & $2.19\times10^{3}$ & $3.20\times10^{27}$ & NUGIS$^{a}$ \\
WR15  & WC6 & $0.33^{*}$ & / & 0.69 & 0.11 & 2500 & 100 & $5.16\times10^{18}$ & $9.21\times10^{17}$ & $5.42\times10^{4}$ & $9.67\times10^{3}$ & $4.44\times10^{28}$ & LEITH$^{c}$ \\
WR93  & WC7 & 0.90 & 0.20 & / & / & 1900 & 100 & $3.89\times10^{18}$ & $9.56\times10^{17}$ & $4.08\times10^{4}$ & $1.00\times10^{4}$ & $1.87\times10^{28}$ & NUGIS$^{a}$ \\
WR135 & WC8 & 0.60 & 0.09 & / & / & 2400 & 150 & $4.14\times10^{18}$ & $8.08\times10^{17}$ & $4.43\times10^{4}$ & $8.48\times10^{3}$ & $1.99\times10^{28}$ & NUGIS$^{a}$ \\
WR81  & WC9 & 0.30 & 0.08 & / & / & 2500 & 100 & $2.24\times10^{18}$ & $6.25\times10^{17}$ & $2.36\times10^{4}$ & $6.56\times10^{3}$ & $1.08\times10^{28}$ & NUGIS$^{a}$ \\
WR6   & WN4 & 1.04 & 0.05 & / & / & 1400 & 100 & $2.44\times10^{18}$ & $3.68\times10^{17}$ & $2.56\times10^{4}$ & $3.86\times10^{3}$ & $1.17\times10^{28}$ & NUGIS$^{a}$ \\
WR141 & WN5 & 0.59 & 0.04 & 1.28 & 0.04 & 1900 & 50 & $2.55\times10^{18}$ & $2.19\times10^{17}$ & $2.68\times10^{4}$ & $2.30\times10^{3}$ & $1.22\times10^{28}$ & MONTES$^{b}$ \\
WR136 & WN6 & 1.95 & 0.09 & / & / & 1700 & 50 & $6.74\times10^{18}$ & $5.04\times10^{17}$ & $7.08\times10^{4}$ & $5.30\times10^{3}$ & $3.24\times10^{28}$ & NUGIS$^{a}$ \\
WR78  & WN7 & 1.50 & 0.09 & / & / & 1600 & 100 & $4.60\times10^{18}$ & $6.37\times10^{17}$ & $4.83\times10^{4}$ & $6.69\times10^{3}$ & $2.21\times10^{28}$ & NUGIS$^{a}$ \\
WR16  & WN8 & 1.21 & 0.09 & 1.75 & 0.09 & 2300 & 100 & $7.66\times10^{18}$ & $8.77\times10^{17}$ & $8.04\times10^{4}$ & $9.20\times10^{3}$ & $3.68\times10^{28}$ & LEITH$^{c}$ \\
\hline
\end{tabular}
\tablefoot{
Table associated with the extrapolation method described in Sect.\,\ref{Section: Estimation_avec_ref}.
Columns list the object name, spectral type, flux densities at 4.8 and 8.6\,GHz and their associated errors, distance and distance errors, spectral luminosities with their errors, $\nu L_{\nu}$, and the reference.
The symbol $^{*}$ indicates an upper limit.
References: $^{a}$ \citet{Nugis1998}, $^{b}$ \citet{Montes2009}, $^{c}$ \citet[]{ClausI,ClausII}.
}
\end{table*}

The adopted references for all available subtypes are listed in Table\,\ref{tab:thermal_references}. For the WC6 subtype, only an upper limit was available at 4.8 GHz; in this case, we used the flux density at 8.6 GHz to compute the spectral luminosity. In addition, a supplementary column is provided showing $\nu L_{\nu}$ (in CGS units), which represents the energy output per unit time.  

To carry out a comparison with the MeerKAT measurements at 1.3\,GHz, the reference luminosities at 4.8 or 8.6\,GHz must be extrapolated to the survey frequency of 1.3\,GHz. However, the actual optically thick thermal spectral index is not known and should lie within a given range, with a lower value of 0.6. Therefore, a median spectral index of 0.7 was adopted, with an uncertainty of 0.1. 

It should be noted that the WN9 subtype was not included in the original references, although it does correspond to some detected objects. For this case, the closest available subtype was used instead WN8, as the stellar wind properties of adjacent subtype should not differ significantly. 
These values are displayed in Table\,\ref{tab:Extrapolation_flux_densities}. The next step therefore consists of associating each detected star with the reference listed in Table\,\ref{tab:Extrapolation_flux_densities} and rescaling the flux densities using the distance to the target, 
\begin{equation}\label{Eq:distance_scale}
    S_{1.3}^{Extra} = S_{1.3}^{D_\mathrm{ref}} \bigg(\frac{D_\mathrm{target}}{D_\mathrm{ref}}\bigg )^2 \, ,
\end{equation}

\noindent where $S_{1.3}^{Extra}$ is the flux density of the reference star rescaled to the distance of the target, $S_{1.3}^{D_\mathrm{ref}}$ is the flux density of the reference star at its actual distance, $d_\mathrm{target}$ is the distance of the target, and ${D_\mathrm{ref}}$ is the distance of the reference star. The uncertainty in the extrapolated flux density is obtained through error propagation of Eq.\,\ref{Eq:distance_scale}. For these objects, the distances come from either \citet{Crowtherdistance}, \citet{CrowtherdistanceGDR2}, or dedicated papers such as \citet{WR76_distance}, \citet{Apep_distance}, and \citet{WR84_distance}. 
Unfortunately, no radio measurements at 4.8 or 8.6 GHz are available for WO stars. Therefore, this method cannot presently be applied to WR30a.

\begin{table}
\caption{Flux densities for the reference stars extrapolated at 1.3\,GHz.}
\label{tab:Extrapolation_flux_densities}
\centering
\begin{tabular}{lccc}
\hline\hline
Sp. type &
Name &
$S_{1.3}^{D_\mathrm{ref}}$ (mJy) &
$\sigma_{S_{1.3}^{D_\mathrm{ref}}}$ (mJy) \\
\hline
WC4 & WR144 & 0.27 & 0.12 \\
WC5 & WR111 & 0.13 & 0.06 \\
WC6 & WR15 & 0.18 & 0.04 \\
WC7 & WR93 & 0.36 & 0.13 \\
WC8 & WR135 & 0.24 & 0.08 \\
WC9 & WR81 & 0.12 & 0.05 \\
\hline
WN4 & WR6 & 0.42 & 0.14 \\
WN5 & WR141 & 0.24 & 0.07 \\
WN6 & WR136 & 0.78 & 0.22 \\
WN7 & WR78 & 0.60 & 0.19 \\
WN8 & WR16 & 0.49 & 0.14 \\
WN9 & WR16$^{\dagger}$ & 0.49 & 0.14 \\
\hline
\end{tabular}
\tablefoot{Since flux density data were not available for the WN9 subtype, the values corresponding to the WN8 subtype were adopted instead; this is indicated by the $\dagger$ symbol.}
\end{table}

\subsection{Search for radio excess}\label{Section:Excess}
The flux density at 1.3 GHz of each detected source ($S_{1.3}$) and their theoretical or predicted estimates are compiled in Table\,\ref{Tab:FluxFull}. To determine whether these sources display a significant radio excess that is potentially attributable to synchrotron radiation, we compared the measured flux densities at 1.3 GHz with the theoretical or extrapolated thermal model sources discussed above.  

To quantify the radio excess, we defined a $Z$-score used as a radio excess significance indicator ($Z$), which is separately related to the two methods described above.
\begin{equation}
Z^\mathrm{Th/Extra} = \frac{\Delta\,S^\mathrm{Th/Extra}}{\sigma} \, ,
\end{equation}
\noindent with
$$\Delta\,S^\mathrm{Th/Extra} = S_{1.3} - S_{1.3}^\mathrm{Th/Extra}\,,$$ 
$$\sigma = \sqrt{\sigma_{1.3}^2 + \big(\sigma_{1.3}^\mathrm{Th/Extra}\big)^2}\,,$$
\noindent where the superscripts 'Th' or 'Extra' refer to the two methods described in Sect.\,\ref{Section:WrightBarlow} and \ref{Section: Estimation_avec_ref}, respectively. In both cases, $\Delta\,S^\mathrm{Th/Extra}$ and $Z^\mathrm{Th/Extra}$ are given in Table\,\ref{Tab:FluxFull}.

Our approach considers three distinct cases:
\begin{itemize}
\item Case A: The target flux density is significantly higher than the thermal estimates ($Z>3$). In this scenario, the radio emission cannot be attributed only to thermal processes from the WR wind. A significant radio excess is measured. These targets are indicated by a flag in Table\,\ref{Tab:FluxFull}.
\item Case B: The target flux density is only marginally greater than the thermal estimate ($1\,\leq\,Z\,\le\,3$). These targets show a potential radio excess, which requires further investigation and follow-up studies to confirm. These targets are also indicated by a flag in Table\,\ref{Tab:FluxFull}.
\item Case C: The measured flux density at 1.3 GHz of the target is comparable to (or even below) the predicted or extrapolated value ($Z\,<\,1$). Thus, there is no radio excess at least at the epoch of the Meerkat measurement. The radio emission could therefore still be explained purely by thermal emission from the WR wind within the uncertainty of the approaches adopted to calculate the thermal emission estimators. 
\end{itemize}

\begin{table*}
\caption{Radio excess diagnostic}
\label{Tab:FluxFull}
\centering
\begin{tabular}{lllcccccccccc}
\hline\hline
\# & Name & Spectral Type & $S_{1.3}$ & $\sigma_{S_{1.3}}$ & $S_{1.3}^{\mathrm{Th}}$ & $\sigma_{S_{1.3}}^{\mathrm{Th}}$ & $Z^{\mathrm{Th}}$ & $S_{1.3}^{\mathrm{Extra}}$ & $\sigma_{S_{1.3}}^{\mathrm{Extra}}$ & $Z^{\mathrm{Extra}}$ & Case & Flag \\
& & & (mJy) & (mJy) & (mJy) & (mJy) & & (mJy) & (mJy) & & & \\
\hline \\
1 & WR 105 & WN9h & 0.43 & 0.15 & 1.18 & 1.16 & -0.65 & 0.21 & 0.06 & 1.31 & C,B & $\checkmark$\ \\
2 & WR 111 & WC5 & 0.14 & 0.06 & 0.54 & 0.53 & -0.75 & 0.13 & 0.06 & 0.12 & C,C & \\
3 & WR 110 & WN5-6b & 0.10 & 0.02 & 0.80 & 0.78 & -0.90 & 0.26 & 0.07 & -2.19 & C,C & \\
4 & WR 112 & WC9d+OB? & 2.35 & 0.04 & 0.02 & 0.02 & 50.11 & 0.01 & 0.01 & 58.26 & A,A & $\checkmark$\ \\
5 & WR 118-3 & WN9 & 0.83 & 0.67 & 7.83 & 7.67 & -0.91 & 3.95 & 1.23 & -2.22 & C,C & \\
6 & WR 122 & WN9 & 0.98 & 0.10 & 0.73 & 0.71 & 0.36 & 0.37 & 0.11 & 4.09 & C,A & $\checkmark$\ \\
7 & WR 124-10 & WC6 & 0.24 & 0.03 & 2.44 & 2.39 & -0.92 & 1.71 & 0.39 & -3.78 & C,C & \\
8 & WR 124-2 & WC8 & 2.70 & 0.72 & 0.09 & 0.09 & 3.58 & 0.07 & 0.03 & 3.63 & A,A & $\checkmark$\ \\
9 & WR 125-3 & WN7 & 1.56 & 0.38 & 0.11 & 0.11 & 3.68 & 0.04 & 0.01 & 4.04 & A,A & $\checkmark$\ \\
10 & WR 15 & WC6 & 0.26 & 0.04 & 0.58 & 0.57 & -0.56 & 0.18 & 0.04 & 1.44 & C,B & $\checkmark$\ \\
11 & WR 30a & WO4+O5-5.5 & 0.25 & 0.05 & 0.01 & 0.01 & 4.49 & / & / & / & A,/ & $\checkmark$\ \\
12 & WR 38b & WC7 & 1.57 & 0.06 & 0.04 & 0.04 & 20.68 & 0.04 & 0.02 & 24.67 & A,A & $\checkmark$\ \\
13 & WR 46-3/4 & WN7-8; O6-7.5If+ & 1.02 & 0.38 & 0.09 & 0.09 & 2.41 & 0.03 & 0.01 & 2.63 & B,B & $\checkmark$\ \\
14 & WR 62 & WN6b & 0.12 & 0.03 & 2.14 & 2.10 & -0.96 & 0.16 & 0.04 & -0.73 & C,C & \\
15 & WR 62-1 & WN7-8h & 12.70 & 0.32 & 0.19 & 0.19 & 33.68 & 0.06 & 0.02 & 39.41 & A,A & $\checkmark$\ \\
16 & WR 70-16 & WC7d+WN & 167.27 & 0.22 & 0.07 & 0.07 & 737.62 & 0.06 & 0.02 & 769.03 & A,A & $\checkmark$\ \\
17 & WR 76-7/8 & WN7-8 & 16.93 & 1.01 & 0.04 & 0.04 & 16.75 & 0.01 & 0.01 & 16.79 & A,A & $\checkmark$\ \\
18 & WR 79 & WC7+O5-8 & 0.49 & 0.03 & 0.69 & 0.68 & -0.30 & 0.64 & 0.25 & -0.59 & C,C & \\
19 & WR 84-11 & WN9h & 0.16 & 0.03 & 0.08 & 0.08 & 1.04 & 0.04 & 0.01 & 3.91 & B,A & $\checkmark$\ \\
20 & WR 84-6/7 & WN8-9 & 2.05 & 0.87 & 0.26 & 0.25 & 1.97 & 0.04 & 0.01 & 2.31 & B,B & $\checkmark$\ \\
21 & WR 84-1 & WN8-9 & 1.47 & 0.43 & 0.26 & 0.25 & 2.40 & 0.04 & 0.01 & 3.28 & B,A & $\checkmark$\ \\
22 & WR 94-3 & WN6+abs & 0.02 & 0.01 & 0.36 & 0.36 & -0.96 & 0.22 & 0.06 & -3.19 & C,C & \\
23 & WR 98 & WN8o/C7 & 0.15 & 0.02 & 3.35 & 3.28 & -0.98 & 0.53 & 0.16 & -2.33 & C,C & \\
\hline
\end{tabular}
\tablefoot{The columns show the name of the star, its spectral type, the measured flux density at 1.3 GHz and associated uncertainty, the theoretically predicted flux density and its uncertainty together with the corresponding Z-score, and the extrapolated flux density and its uncertainty together with the corresponding Z-score. The last two columns indicate the classification case associated with each method and a flag if the target falls into Case A or B.}
\end{table*}

We find that the prediction is lower than the measurement for more than one third of the detected targets. This is in agreement with the nature of faint WR stars, which produce less radio emission than predicted by the theory by \citep{WrightBarlow}, as also reported by previous studies \citep[see e.g.][]{Saha2023,Blanco2024}. This highlights significant limitations of the theory in accurately reproducing the observed thermal radio emission from WR stellar winds. Regarding the extrapolated method, the number of targets presenting a negative $Z$ score is lower. The empirical approach is not affected by the limitations of theoretical predictions but is unavoidably affected by other caveats. Typically, the scaling of the extrapolated thermal flux relies on the assumption that the reference stars quoted in Table\,\ref{tab:Extrapolation_flux_densities} are representative of the targets to which they are compared. This illustrates the real difficulty of inferring consistently quantitative values for the thermal emission from WR stars, which complicates the discussion of potential radio excesses.

Given these difficulties, our conservative decision is to report on a radio emission excess for any target corresponding to case A for at least one of the two approaches adopted in this study (11 targets). It has been claimed that targets corresponding to case B from at least one method (four  targets) display a potential radio excess. Case C does not show any excess and is not considered further in our discussion. As a result, 15 detected WR stars out of 23 detections deserve some specific attention.

\subsection{New PACWB candidates}
The main result of Sect.\,\ref{Section:Excess} is the identification of a series of WR stars with a potential or significant excess of radio emission compared to our estimators of thermal emission from the wind. Before identifying these radio-excess emitters as potential synchrotron sources, we must first consider likely alternative physical circumstances that might produce radio emission on top of the thermal wind emission, without any connection to synchrotron radiation produced in a colliding-wind region. 

First, several evolved massive stars, including WR and luminous blue variables, are known to produce a wind-blown bubble accompanied by significant shell-like circumstellar radio emission \citep[e.g.][]{Duncan2002,Cappa2006}. In these cases, the emission can result from a combination of free-free emission from the ionised part of the shell and neutral hydrogen spin-flip transition at 21 cm in the cooled-down part of the shell. Across the typical parameter space of stellar winds of massive stars, the shell radius is typically a few pc, or a few tens of pc, depending on the wind strength, the age of the bubble, and the ISM density surrounding the star. Assuming a lower limit case with a full size of 1 pc \citep{Esteban1994,StockBarlow2010}, at the resolution of the SMGPS, where all our detected sources are located (i.e. 8 arcseconds), circumstellar shells  would reveal their ring-like morphology up to a distance of about 26\,kpc. Even our most distant detected sources, at about 10 kpc, would thus be expected to reveal their extended structure. The fact that all the targets discussed in Sect.\,\ref{Section:Excess} are point-like completely rules out thermal circumstellar shells as the origin of the radio excess. In addition to the thermal emission from wind-blown bubbles, a couple of WR stars appear to be associated with synchrotron emission, likely due to relativistic electrons accelerated in circumstellar shocks \citep{Prajapati2019,Saha2026}. Here, again, this scenario is ruled out by the size of the circumstellar structure, which is typically determined by considerations similar to those for the thermal shells discussed above. Once again, such non-thermal circumstellar emission would be resolved by MeerKAT observations, rejecting the hypothesis of their involvement in the detected radio excess. One last scenario to consider is that of a young supernova remnant positionally coincident with the WR star of interest, either as a line-of-sight occurrence or resulting from a supernova explosion in the direct vicinity of the WR star. The latter case could occur in a binary system where the investigated WR star is the remnant of a binary with a formerly more massive companion. At the end of the free expansion phase of a supernova remnant, which typically lasts a few centuries, a shell radius of a few pc is already developed. This means that a small supernova remnant would be very young, typically at most 100\,yr for a shell radius of 1 pc. At the angular scale of the SMGPS resolution, such a shell would be much smaller than 1\,pc, given the distance to the WR stars considered in Sect.\,\ref{Section:Excess}. As a result, such a small-sized supernova remnant shell would have resulted from a supernova that occurred (at most) a few decades ago, which is highly unlikely. We therefore conclude that such a small-sized supernova remnant scenario cannot be seen as a valid candidate to explain the presence of the radio emission excess we detected. According to the considerations addressed above, the 15 objects displaying a potential or significant radio excess cannot be associated with any case of circumstellar radio emission explained by the interaction of a massive star with the surrounding interstellar medium, regardless of its stage of evolution.

Among the objects that display some radio excess in Table\,\ref{Tab:FluxFull}, three are already known to be PACWBs (see Sect\,.\ref{Sect: knownPACWB}) : WR\,105, WR\,112, and WR\,70-16 (Apep). As a result, the list of targets of interest can be reduced accordingly. We define new PACWB candidates as any target that (i) displays a radio excess; (ii) whose radio emission is unlikely to be explained by unresolved circumstellar emission; and (iii) is not an already established PACWB. The final list of new candidates is thereby reduced to 12 objects (8 Case A and 4 Case B objects, according to the criteria established in Sect.\,\ref{Section:Excess}).

\section{Conclusions}\label{Section: Conclusions}
In this paper, we present a census of 1.3-GHz radio measurements for 429 WR stars based on the SMGPS and MGCS surveys. These stars were compiled into a catalogue including all known WRs located within the longitude and latitude ranges covered by the SMGPS and the MGCS. Among them, 23 were detected and assigned a flux density, while upper limits are provided for 279 non-detected sources. A total of 124 WR stars are located in a region of high radio background, preventing any relevant flux density or upper limit determination. Considering only WR stars located in regions of low background emission, the detection statistics result in a detection rate of about 7.6\%. Such a low detection rate may either be explained by a lack of intrinsic synchrotron radio emission (for single stars or in cases where particle acceleration is lacking) or by the occurrence of strong FFA, which is a dominant absorption process for synchrotron radiation in PACWBs, especially at the frequency of the MeerKAT surveys considered here. 

The main motivation of this work is to characterise the nature of the radio emission of the detected sources, whether they are thermal or non-thermal. For this purpose, we searched for some radio excess by comparing flux measurements to estimators of the thermal free-free emission from WR star winds. The thermal emission estimators were determined using two different methods. The first method consists of a theoretical prediction of the free-free emission using the standard isothermal and isovelocity wind model commonly employed for massive stars, while the second method is based on the extrapolation of radio flux densities measured at higher frequencies for stars of the same spectral type as the target, which display pure thermal behaviour. Both techniques suffer from independent limitations: the thermal prediction fails to adequately reproduce the thermal emission from WR star winds; furthermore, the extrapolation explicitly assumes that the target stars behave like the reference object of the same type. For these reasons, we only claim that the extrapolated fluxes are estimators of the thermal emission level at 1.3\,GHz, and not absolute determinations. Using a Z-score criterion, 11 targets were classified as exhibiting a significant excess compared to the thermal estimators and 4 as exhibiting a potential excess. Based on basic considerations about the potential circumstellar emission that might arise in the vicinity of WR stars, we ruled out the possibility that the measured excesses could originate from circumstellar shells of wind-blown bubbles or non-thermal nebulae. After removing the three already known PACWBs from our list of detected sources presenting a (significant or potential) excess, 12 sources are finally tagged as new PACWB candidates. Dedicated follow-up observations will be necessary to ascertain the nature of the radio emission and potentially confirm the PACWB nature of these objects. Besides multi-band measurements allowing us to determine spectral indices to better infer the nature of the sources, this will also include better angular resolution imaging. At the resolution of the MeerKAT surveys, crowded star clusters harbour potentially several objects within the synthesised beam. The full association of the detected radio sources with the quoted WR stars deserves to be confirmed as well.

As a final remark, it is important to note that the approach adopted in this study relies on snapshot observations of our targets. Given the variability of synchrotron radiation and free-free absorption in PACWBs, a lack of measured radio excess may result from a low amount of synchrotron radiation escaping the system at the time of observation. This is especially emphasised by the few cases of already known PACWBs that are marginally (or not) detected in the data of MeerKAT surveys. Our census of potential candidates is certainly underestimated as a result, emphasising further the need for multi-epoch campaigns to unveil potential particle accelerators. 

\section{Data availability}
The full version of Table 1 is available at the CDS via
\url{https://cdsarc.cds.unistra.fr/viz-bin/cat/J/A+A/713/A154}.  

\begin{acknowledgements}
The authors would like to thank the referee for the positive report and for the constructive comments that helped improve the manuscript. This research is part of the PANTERA-Stars collaboration, an initiative aimed at fostering research activities on the topic of particle acceleration associated with stellar sources\footnote{\url{https://www.astro.uliege.be/~debecker/pantera}}. The MeerKAT telescope is operated by the South African Radio Astronomy Observatory
(SARAO), which is a facility of the National Research Foundation, an agency of the Department of Science and Innovation. This research has made use of NASA's Astrophysics Data System Bibliographic Services. This publication benefits from the support of the Wallonia-Brussels Federation (Belgium) in the context of the FRIA Doctoral Grant awarded to MT.
\end{acknowledgements}

\bibliographystyle{aa}
\bibliography{bibfile}

\onecolumn
\begin{appendix}
\section{Radio images of the 23 detected sources}
\begin{figure*}[ht!]

\begin{tabular}{ccc}
\includegraphics[width=0.32\textwidth]{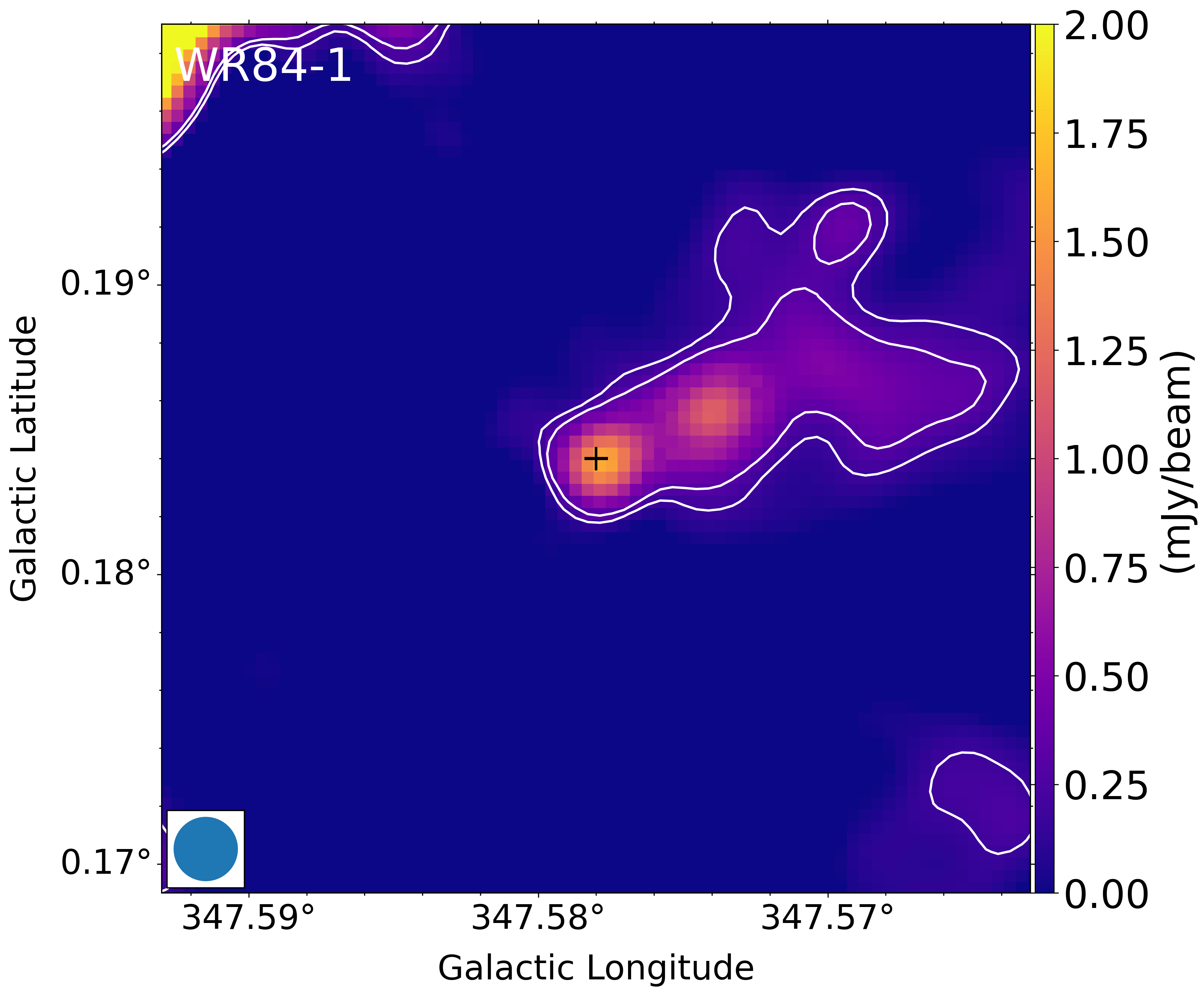} &
\includegraphics[width=0.32\textwidth]{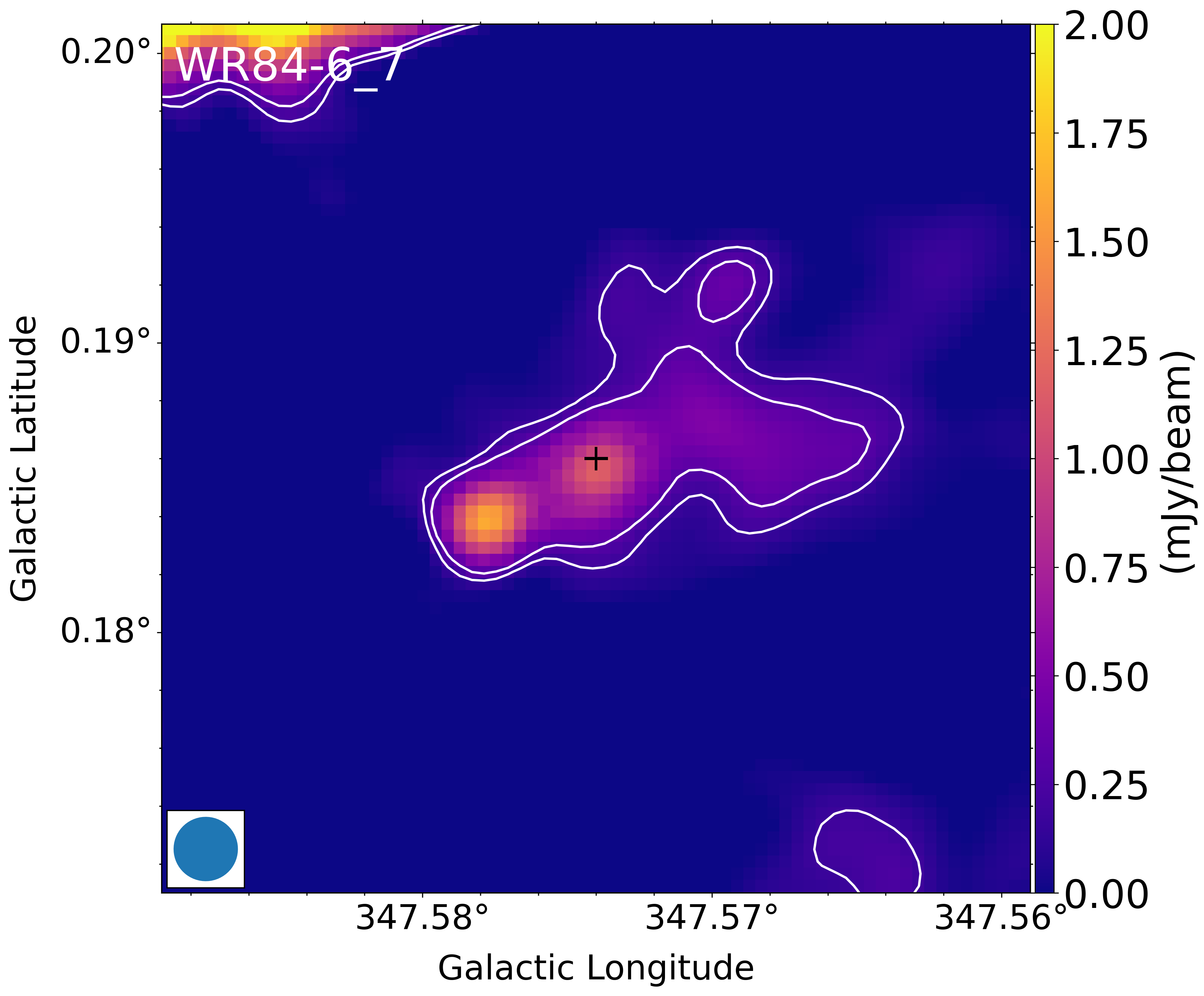} &
\includegraphics[width=0.32\textwidth]{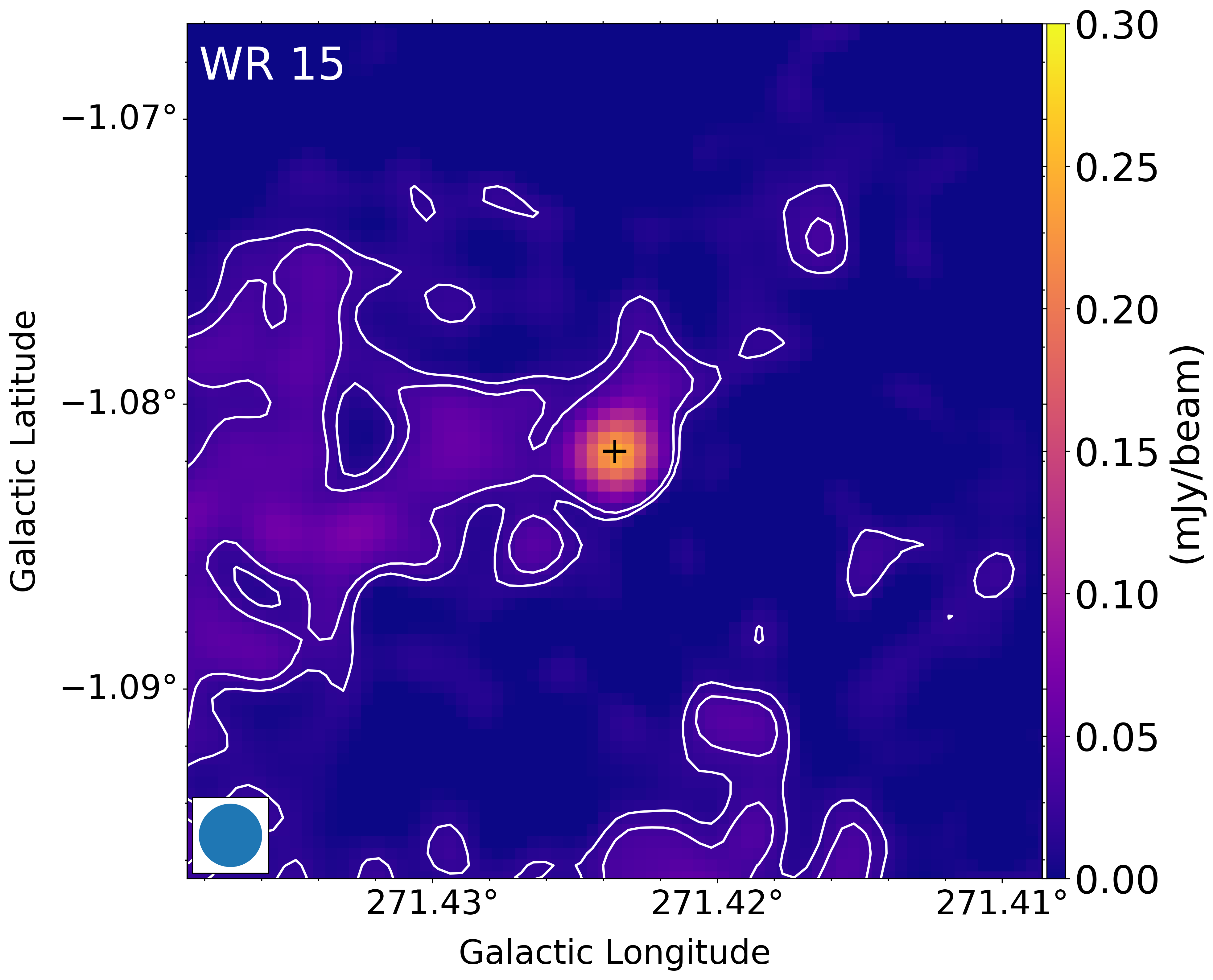} \\

\includegraphics[width=0.32\textwidth]{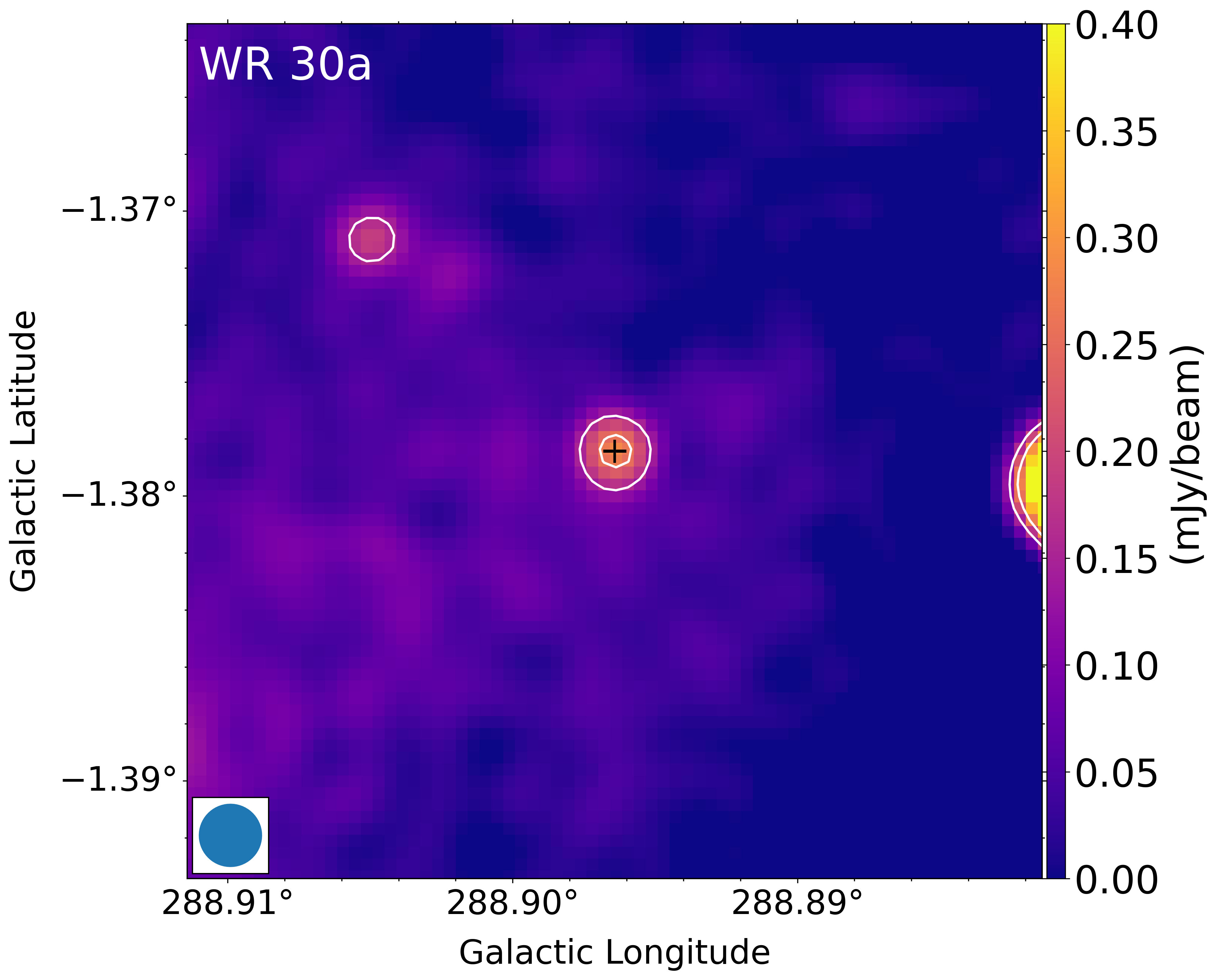} &
\includegraphics[width=0.32\textwidth]{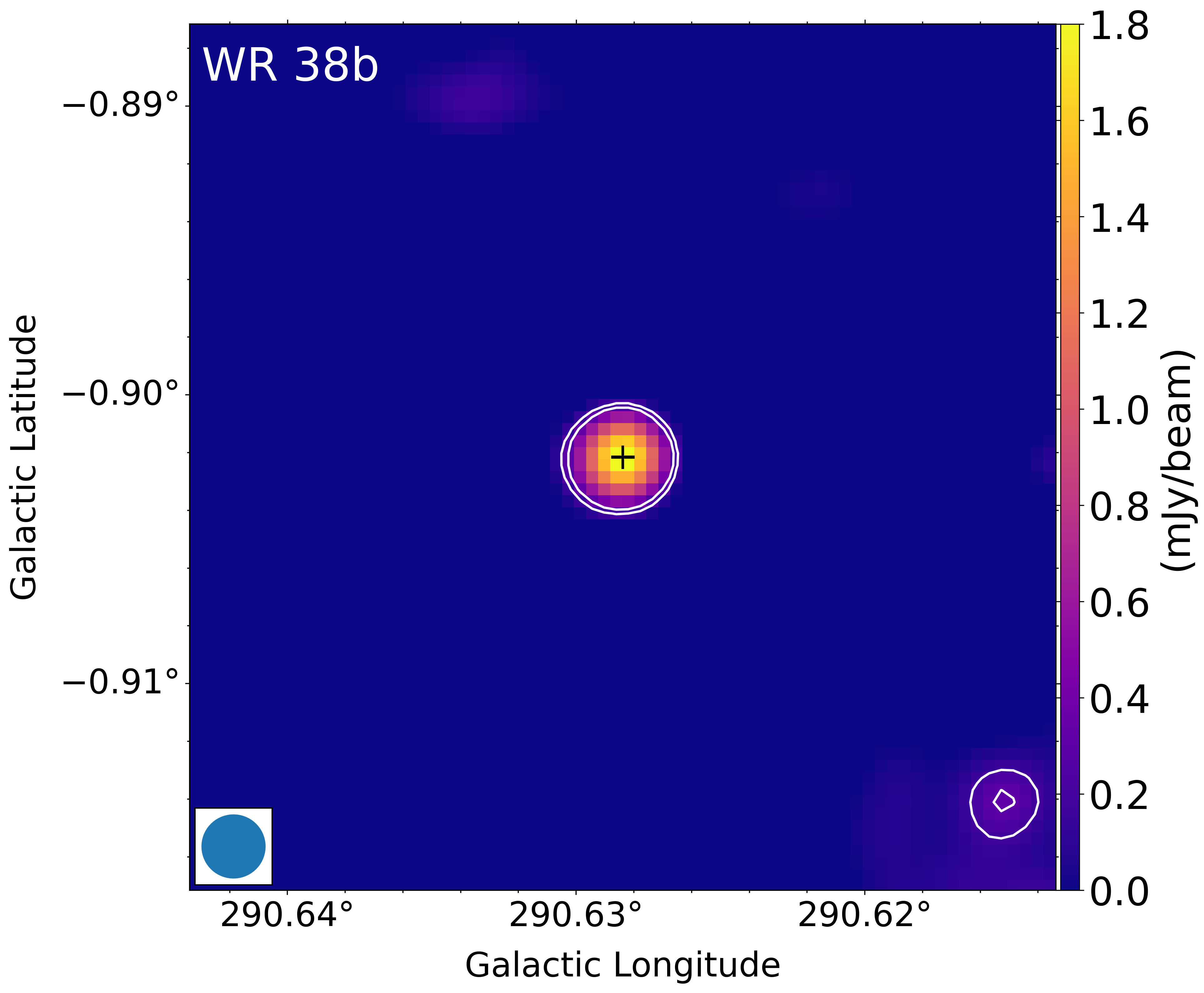} &
\includegraphics[width=0.32\textwidth]{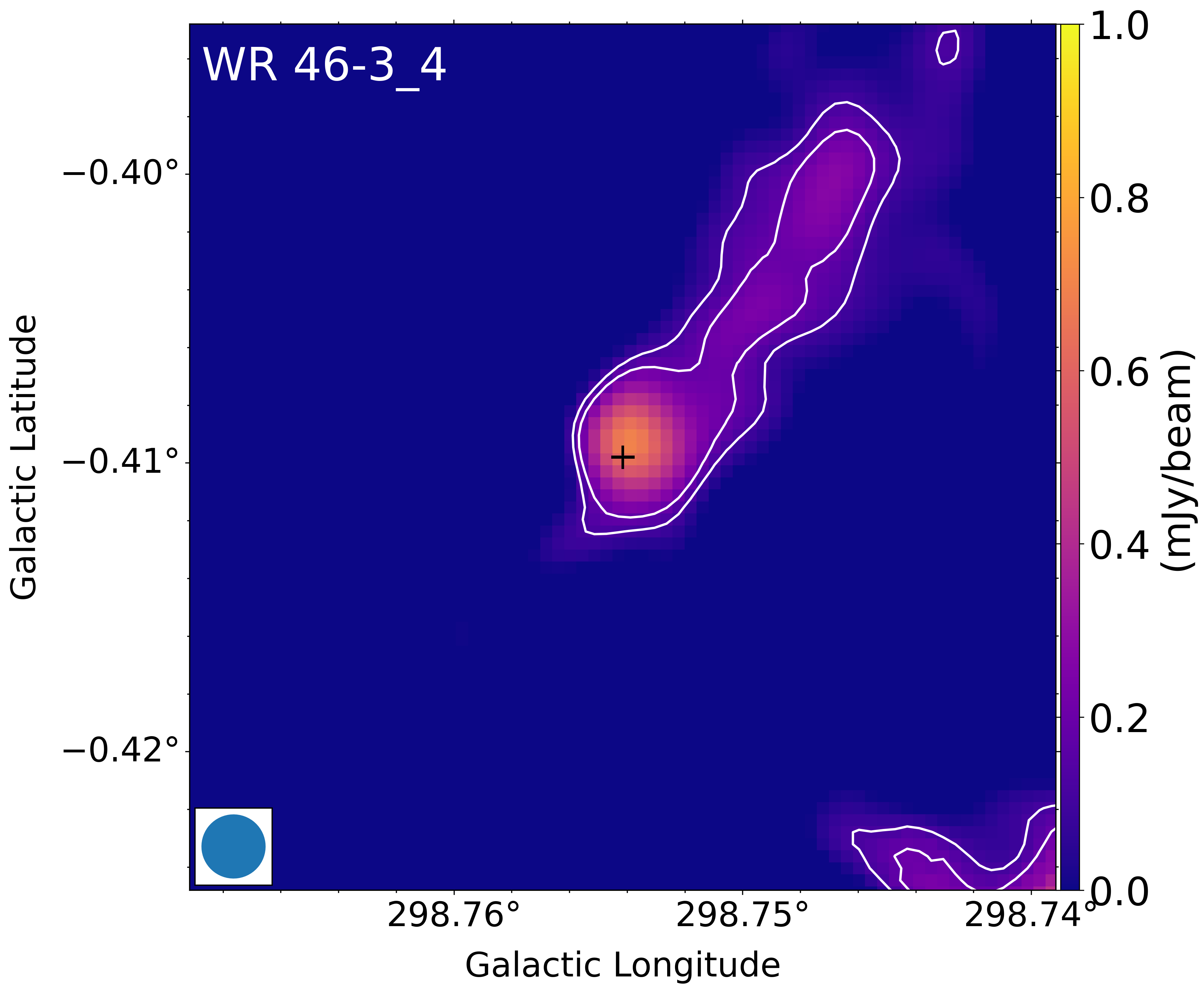} \\

\includegraphics[width=0.32\textwidth]{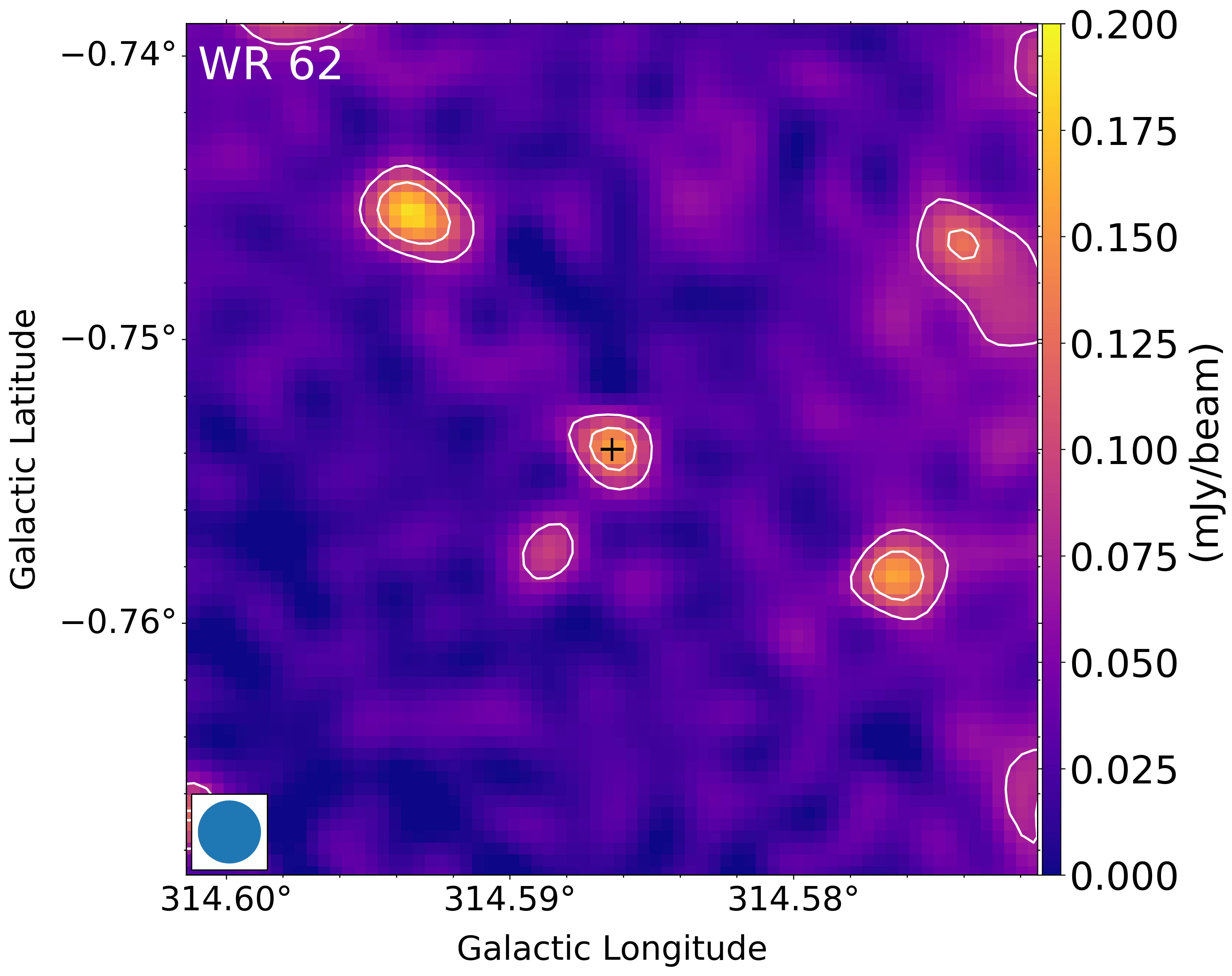} &
\includegraphics[width=0.32\textwidth]{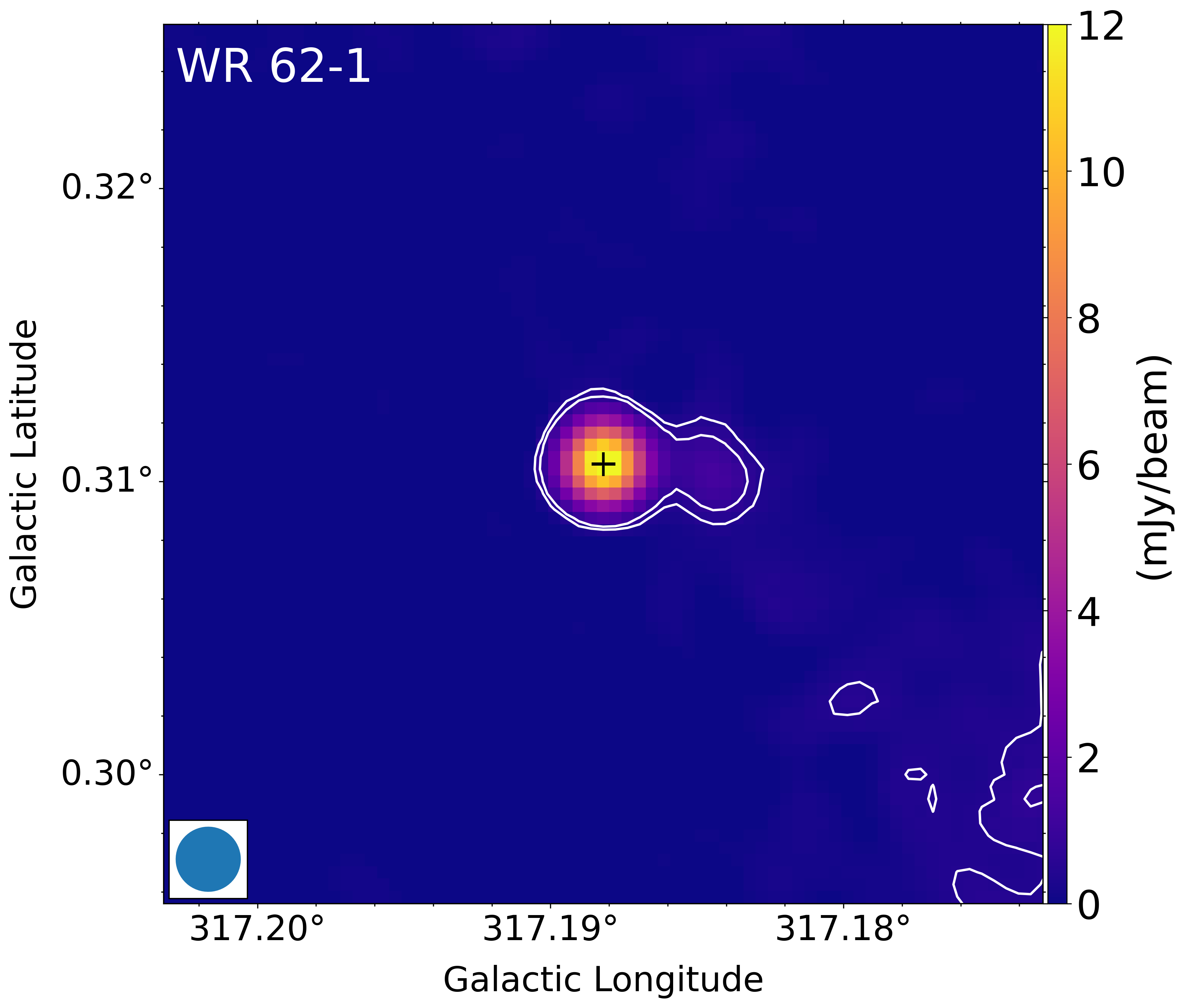} &
\includegraphics[width=0.32\textwidth]{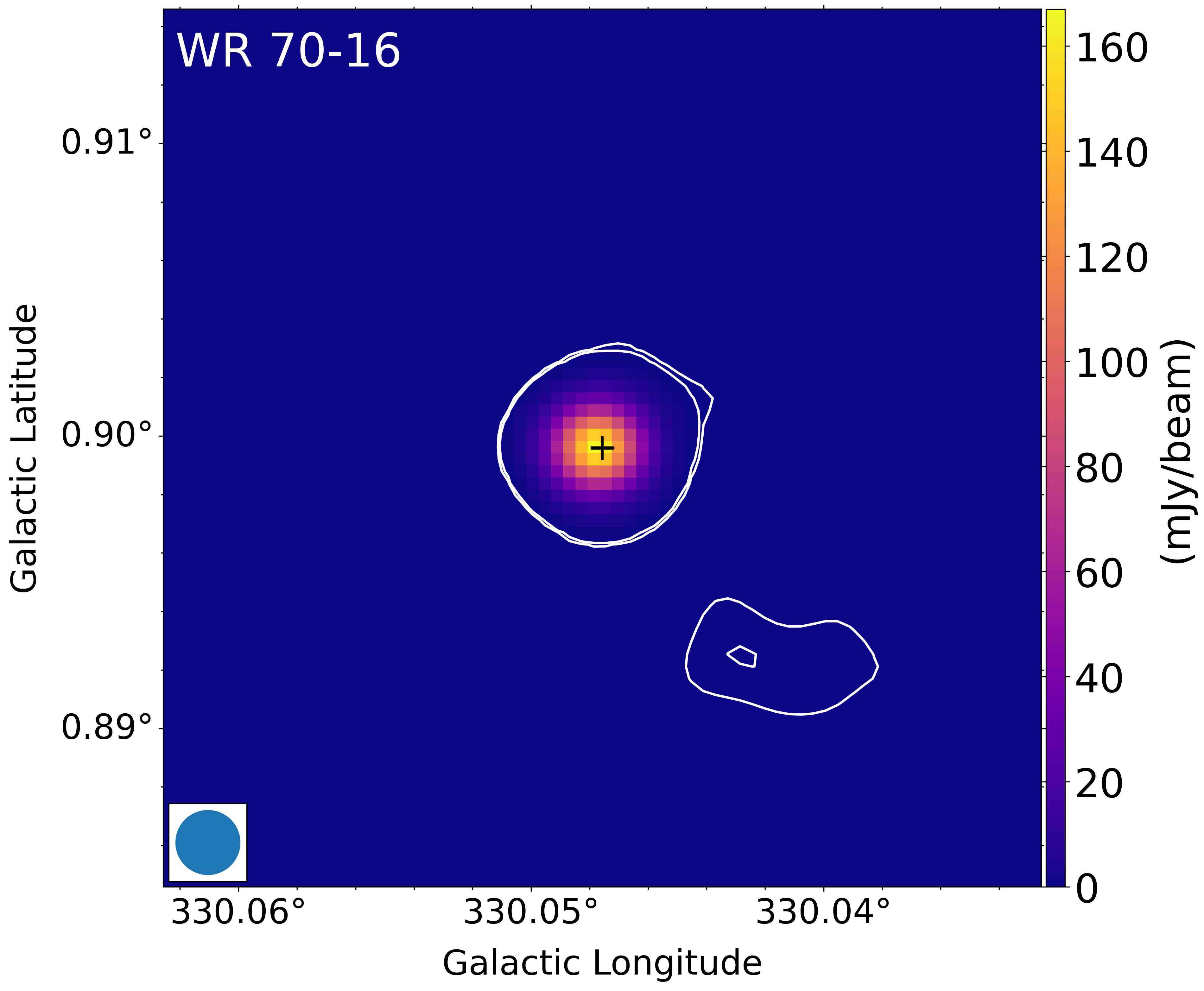} \\

\includegraphics[width=0.32\textwidth]{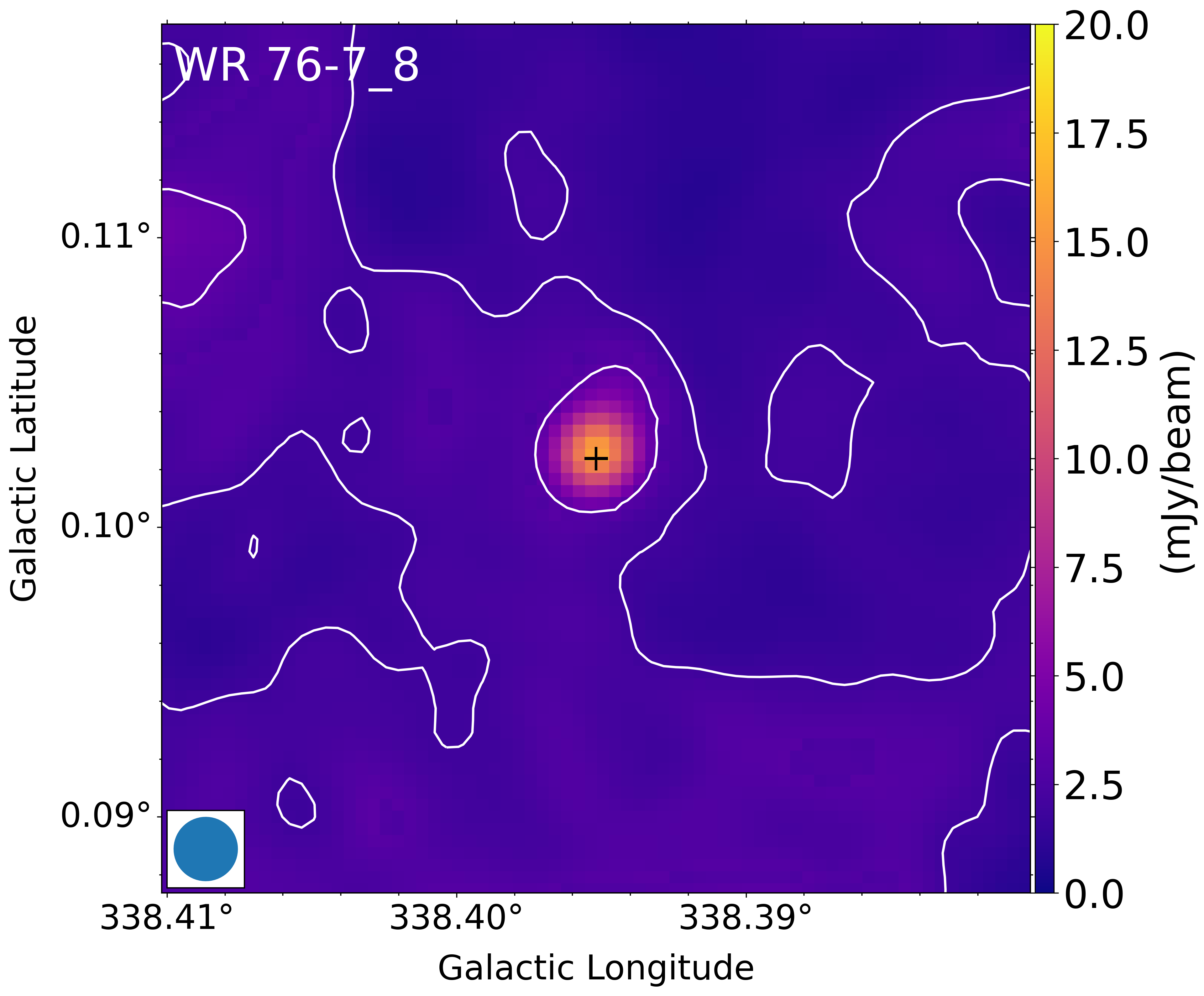} &
\includegraphics[width=0.32\textwidth]{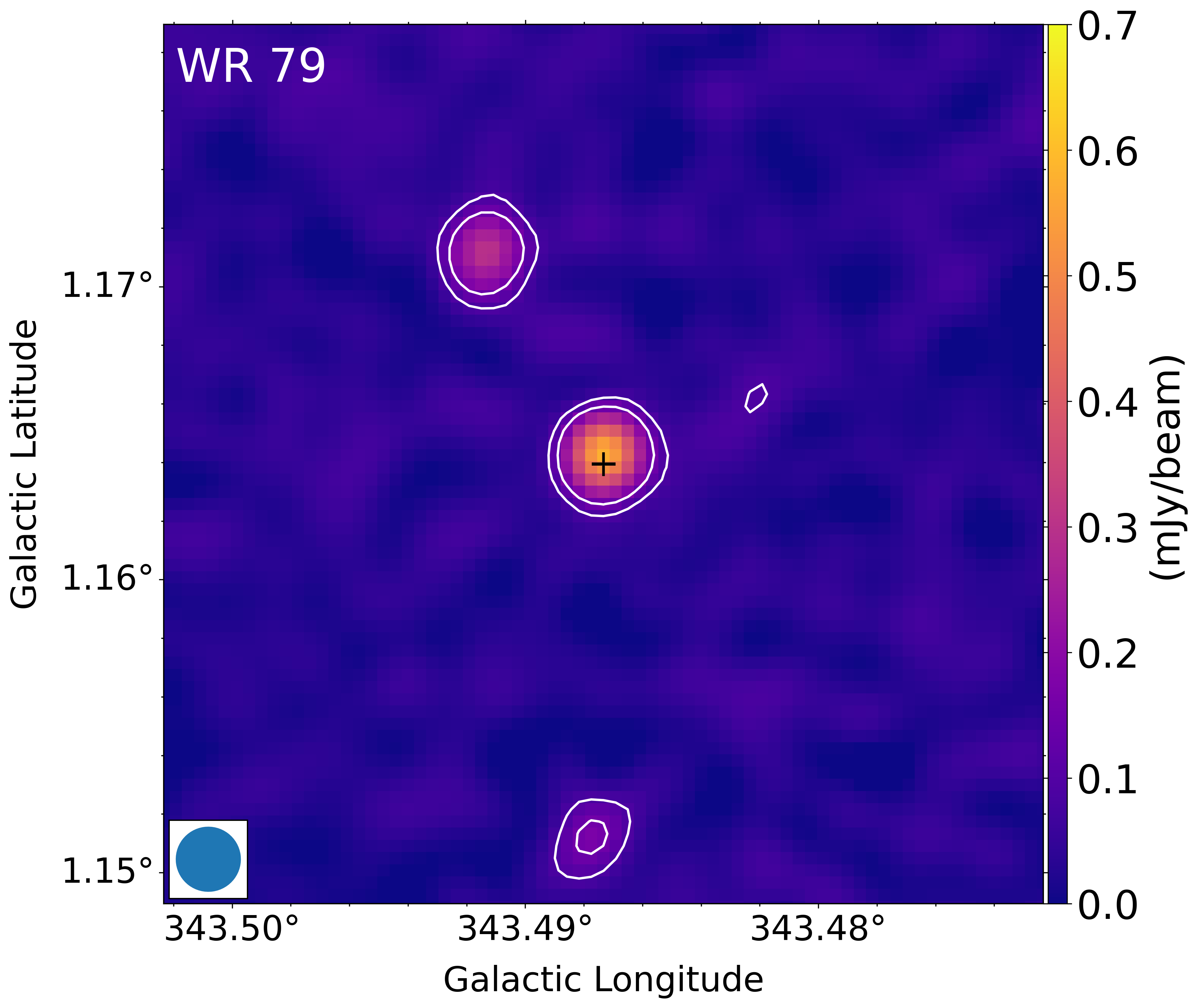} &
\includegraphics[width=0.32\textwidth]{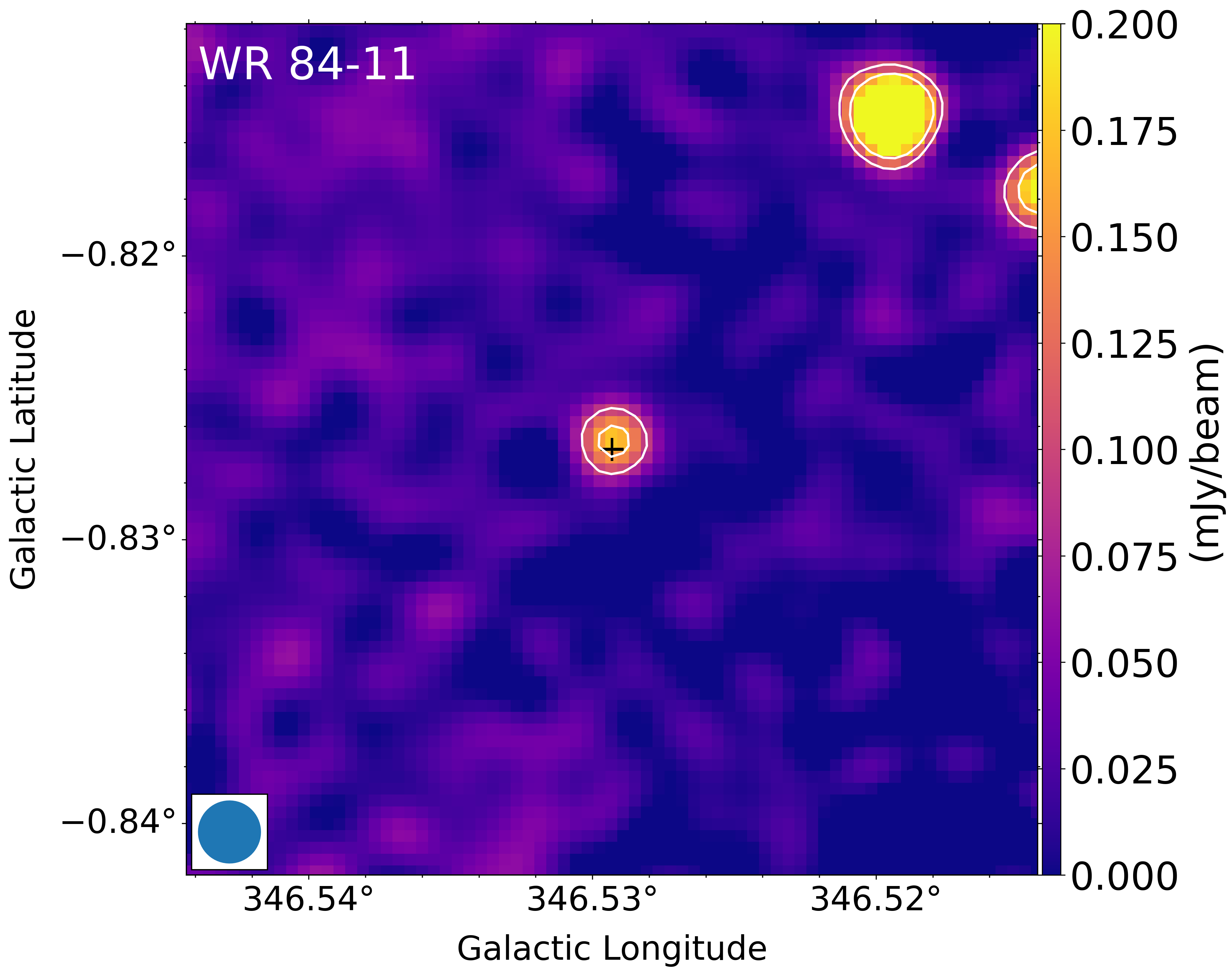} \\
\end{tabular}

\caption{Each source is centered in its respective panel and marked by a cross. The contours correspond to 3 and 5\,$\sigma$ levels. The color scale shows flux density per synthesized beam. The source name is indicated in the top left corner of each panel. The synthesized beam is shown in blue in the bottom left corner of each panel.}
\label{fig:radio_images}
\end{figure*}
\begin{figure*}[ht!]
\ContinuedFloat
\centering
\setlength{\tabcolsep}{2pt}

\begin{tabular}{ccc}
\includegraphics[width=0.32\textwidth]{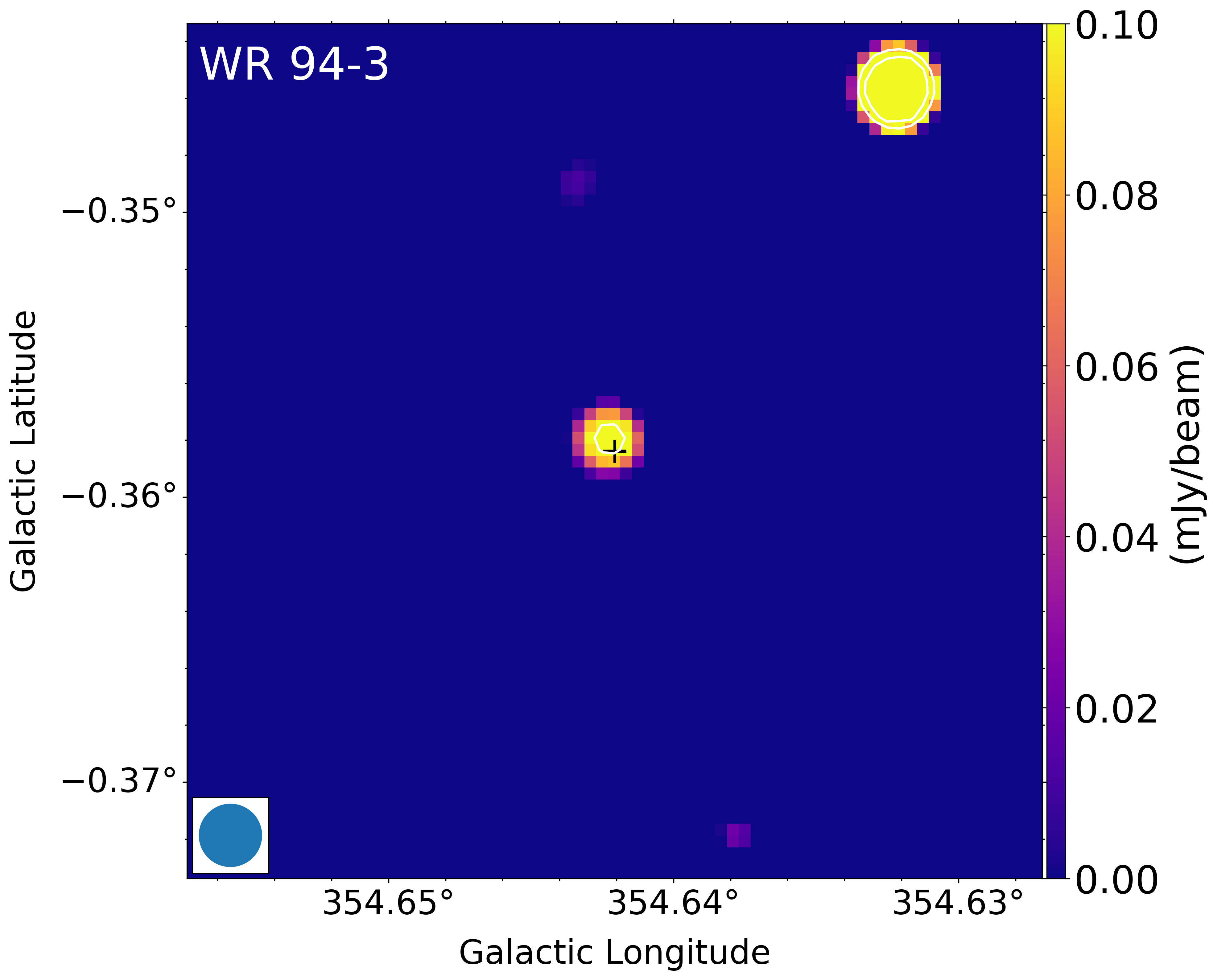} &
\includegraphics[width=0.32\textwidth]{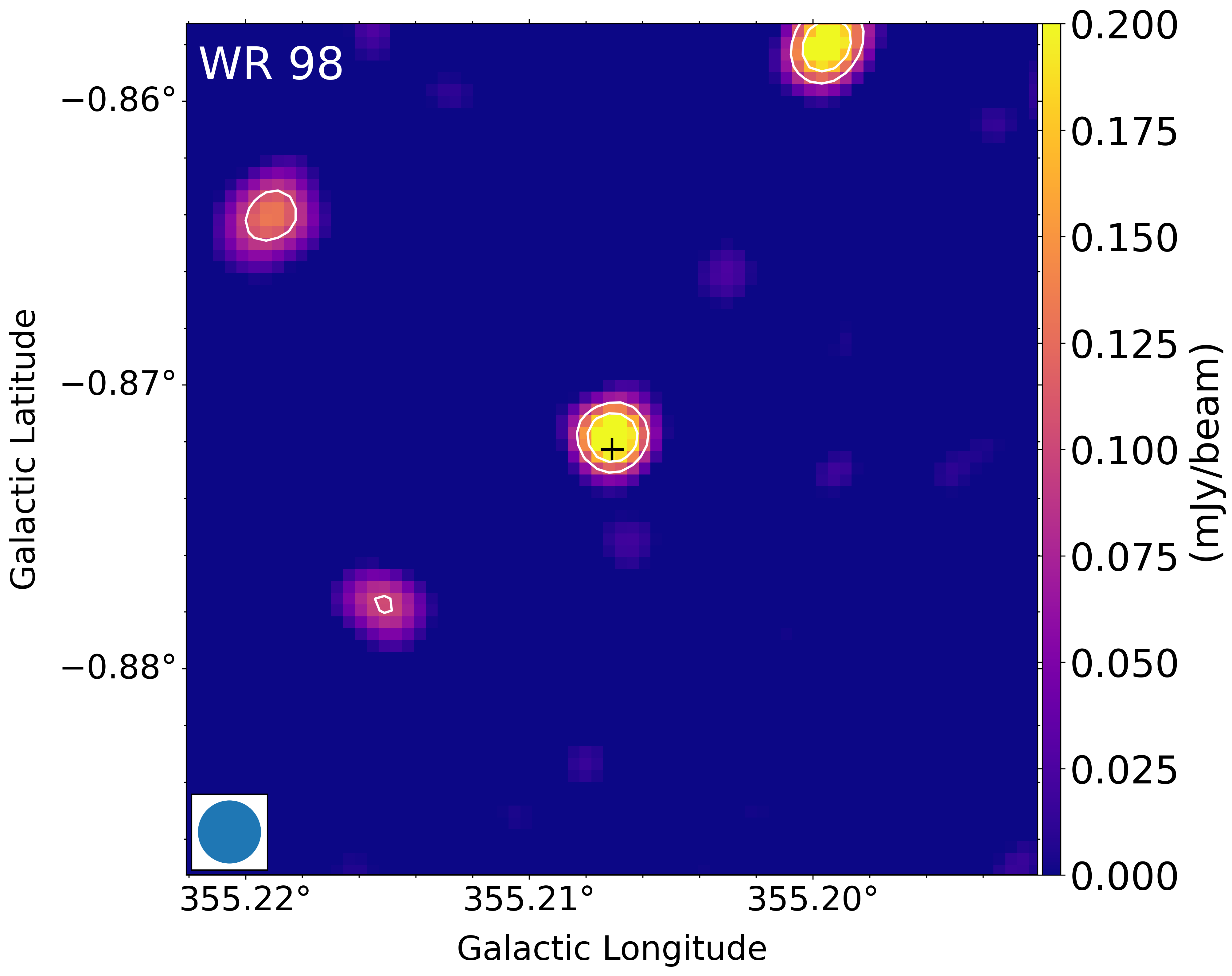} &
\includegraphics[width=0.32\textwidth]{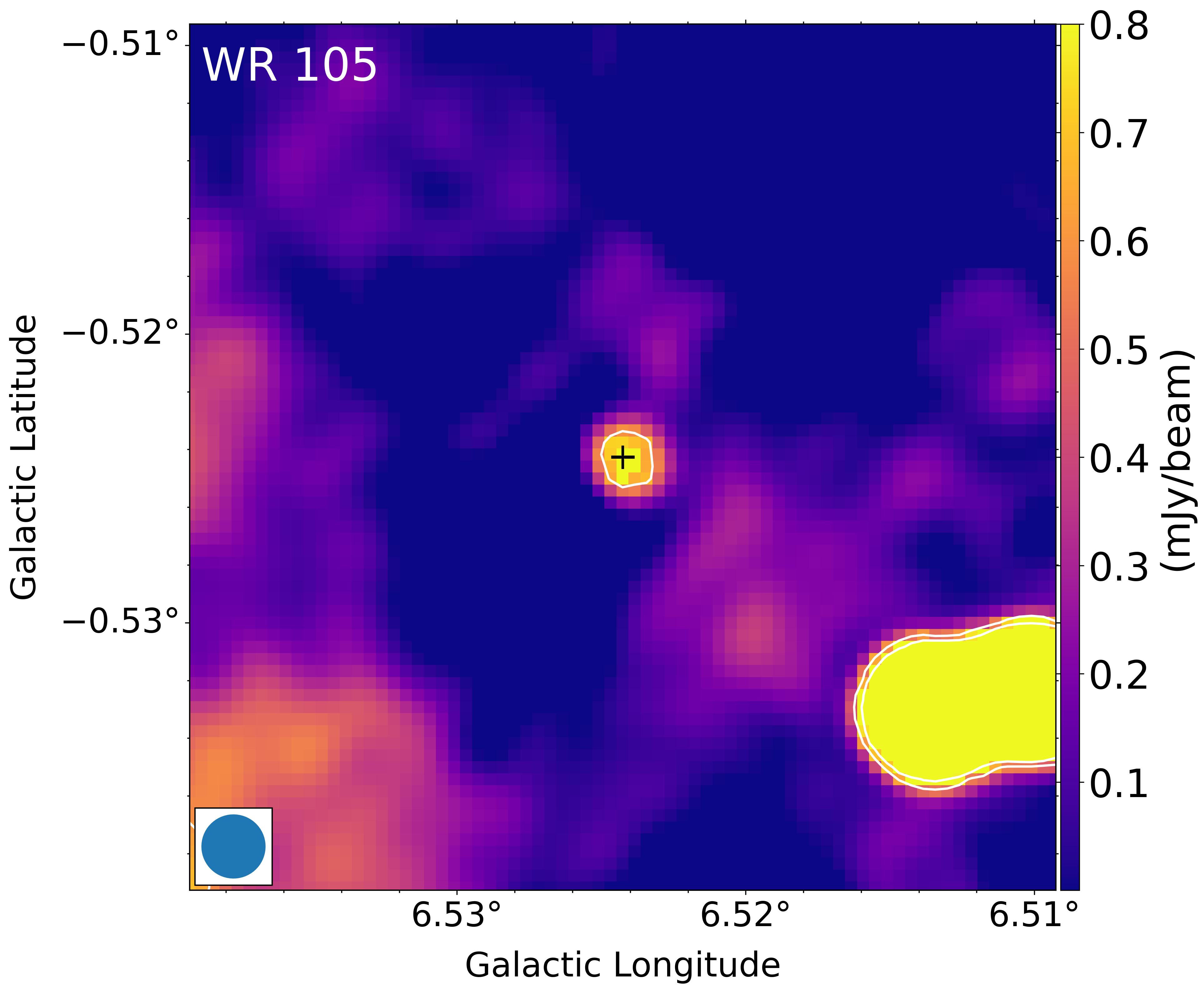} \\

\includegraphics[width=0.32\textwidth]{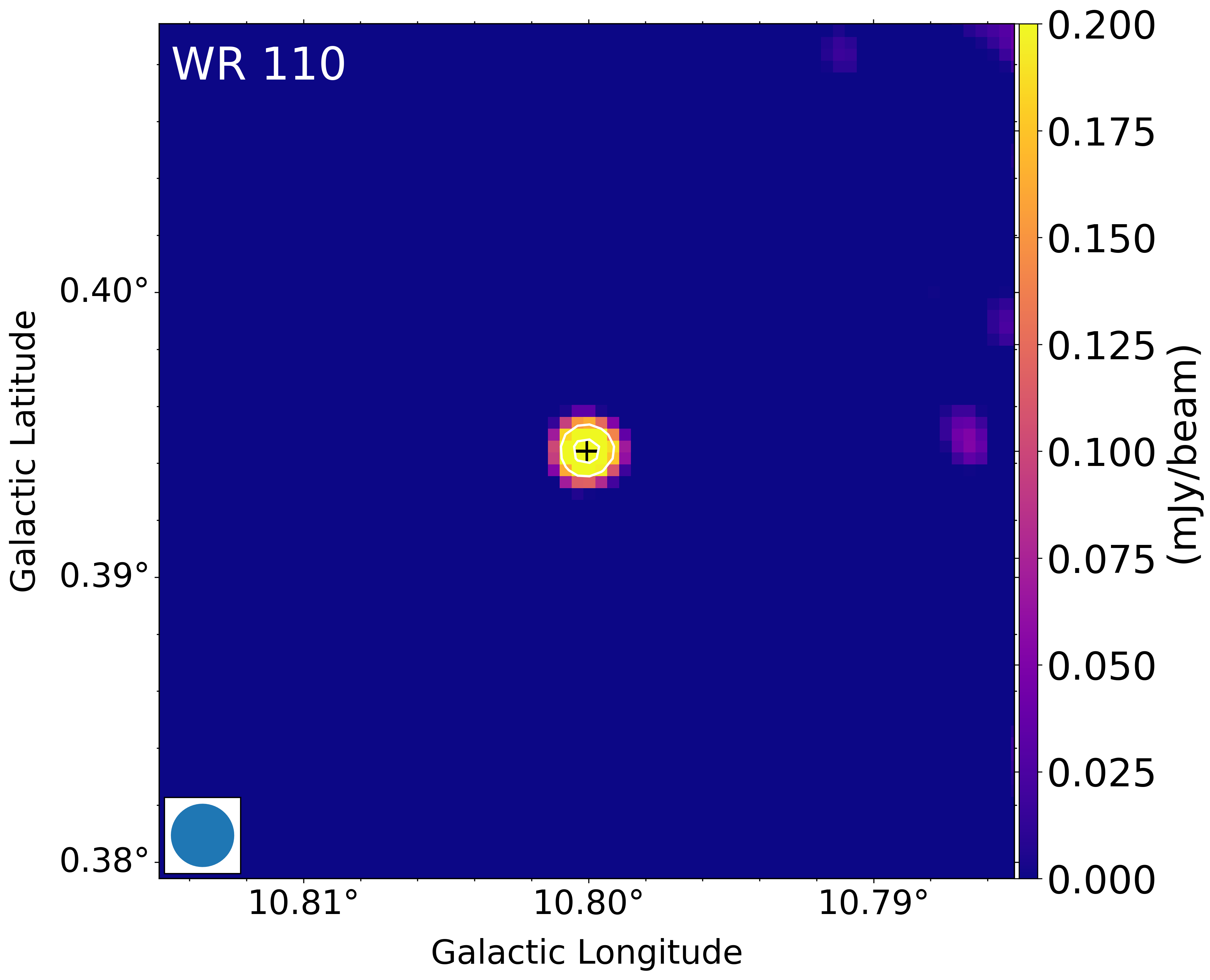} &
\includegraphics[width=0.32\textwidth]{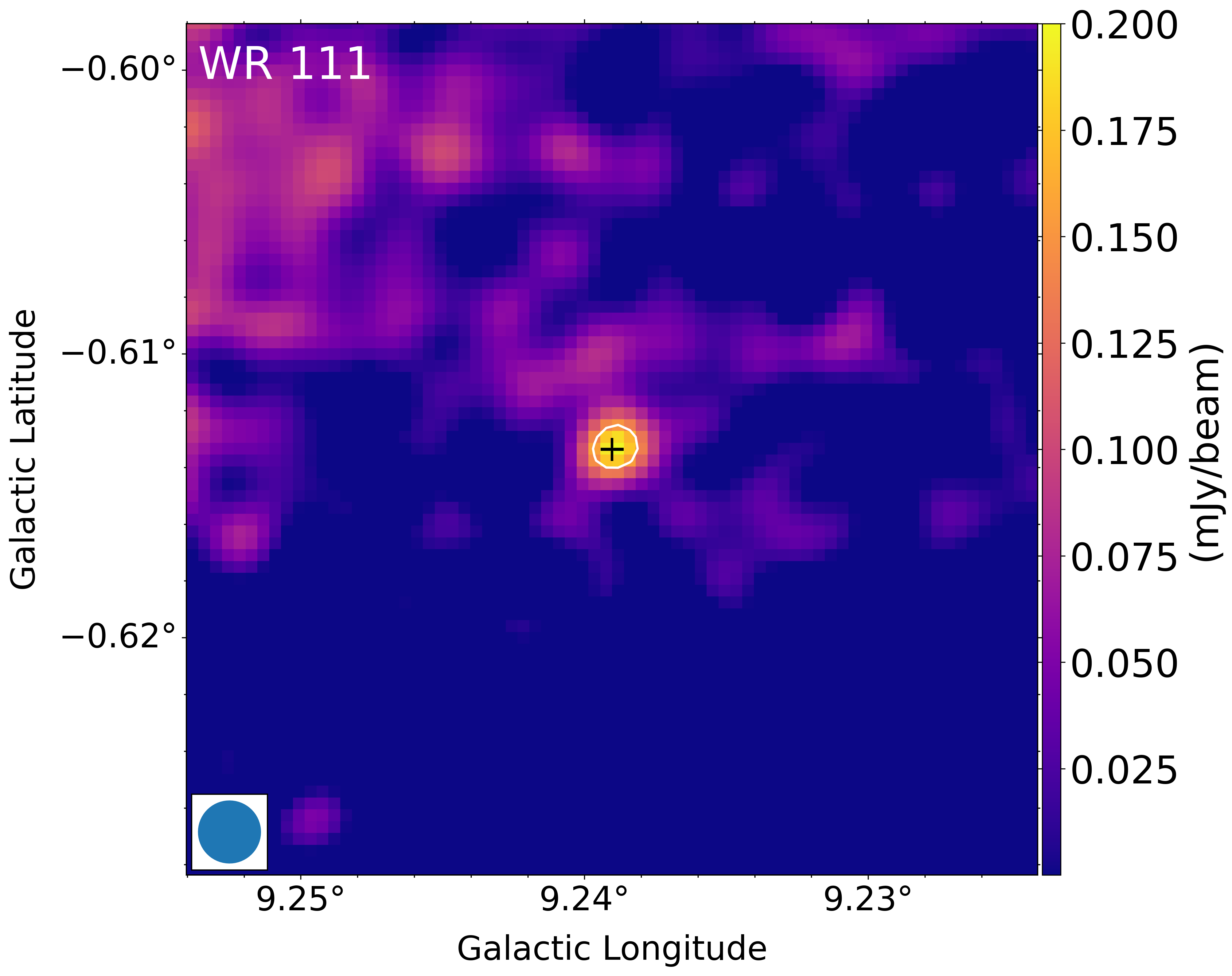} &
\includegraphics[width=0.32\textwidth]{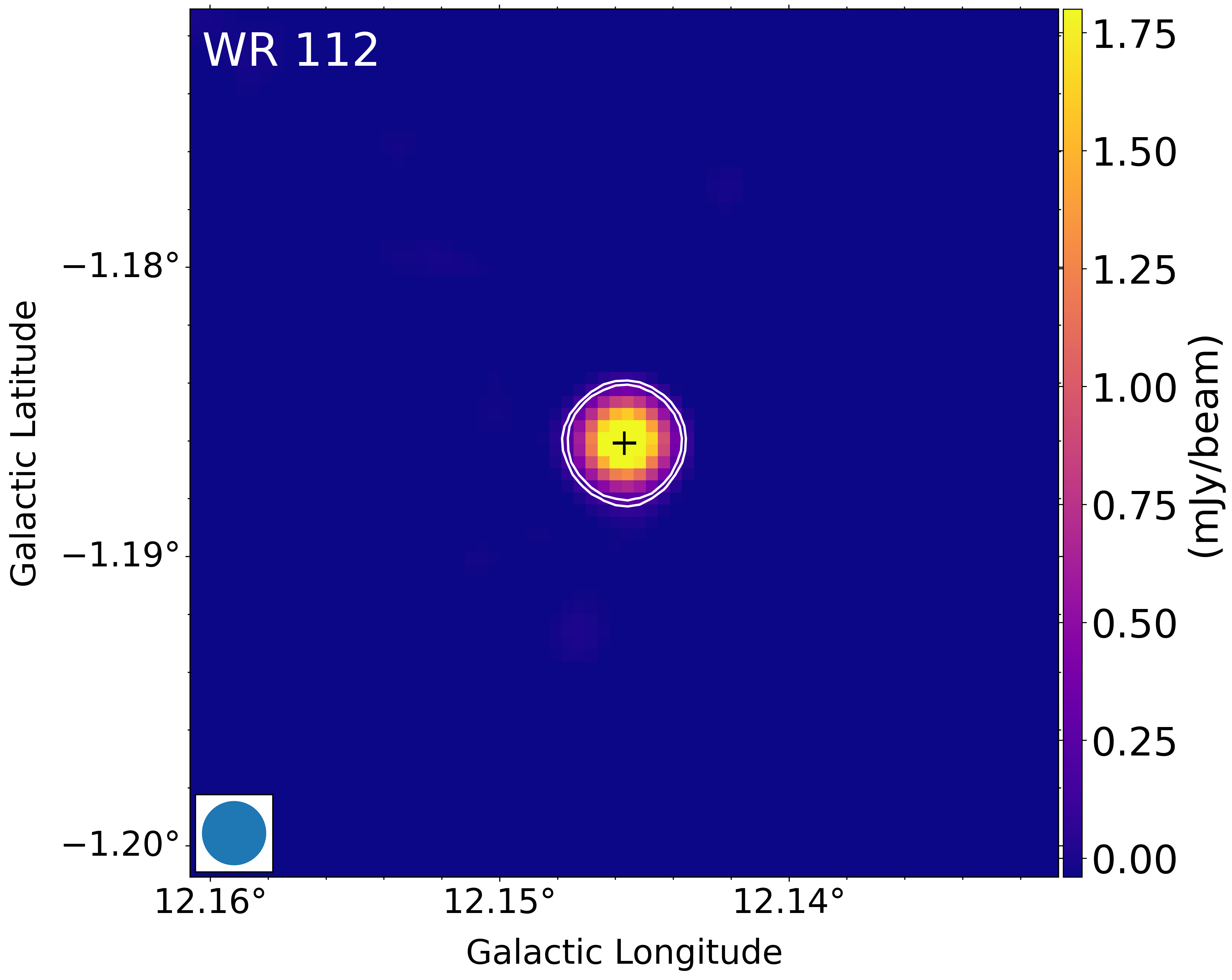} \\

\includegraphics[width=0.32\textwidth]{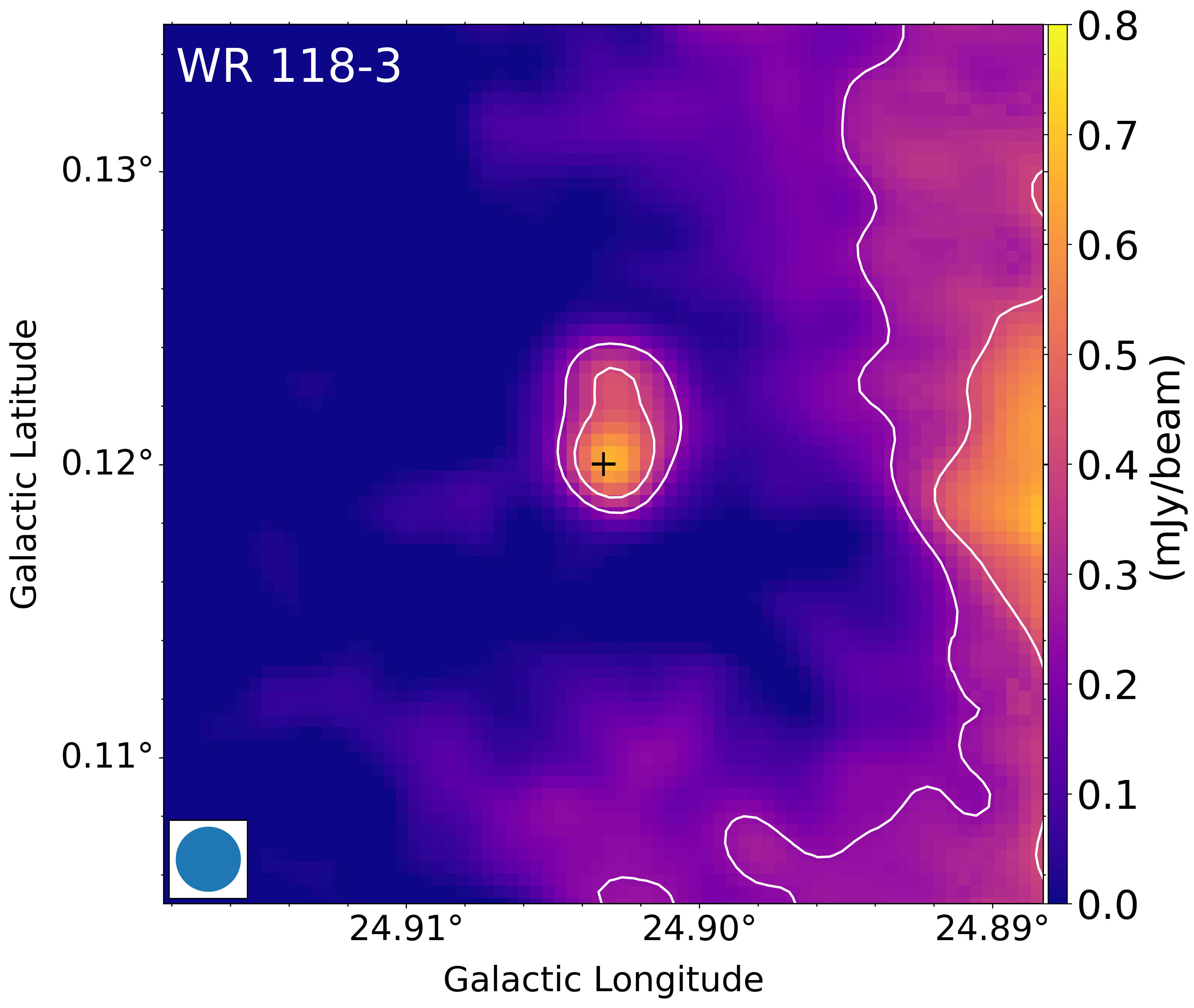} &
\includegraphics[width=0.32\textwidth]{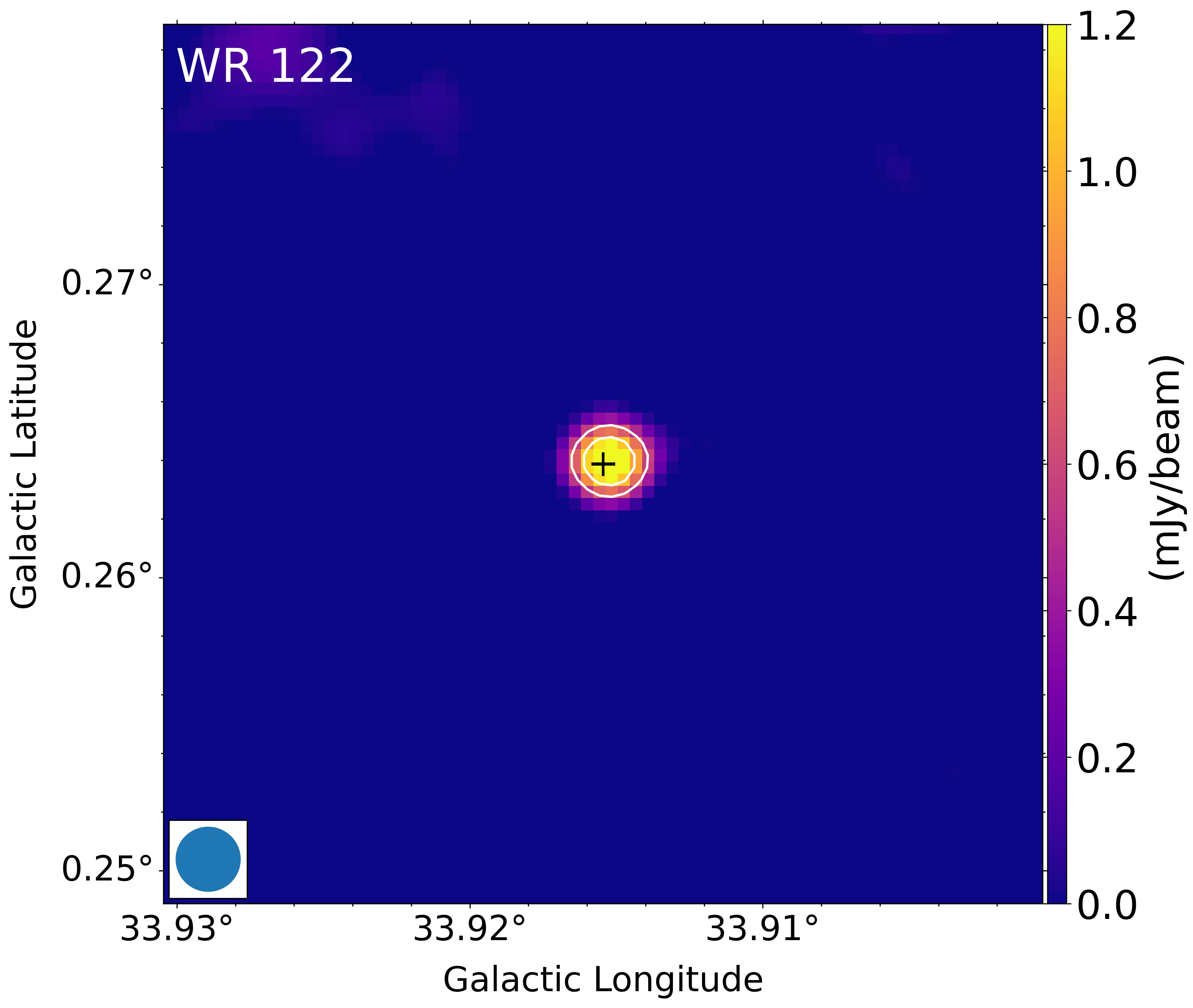} &
\includegraphics[width=0.32\textwidth]{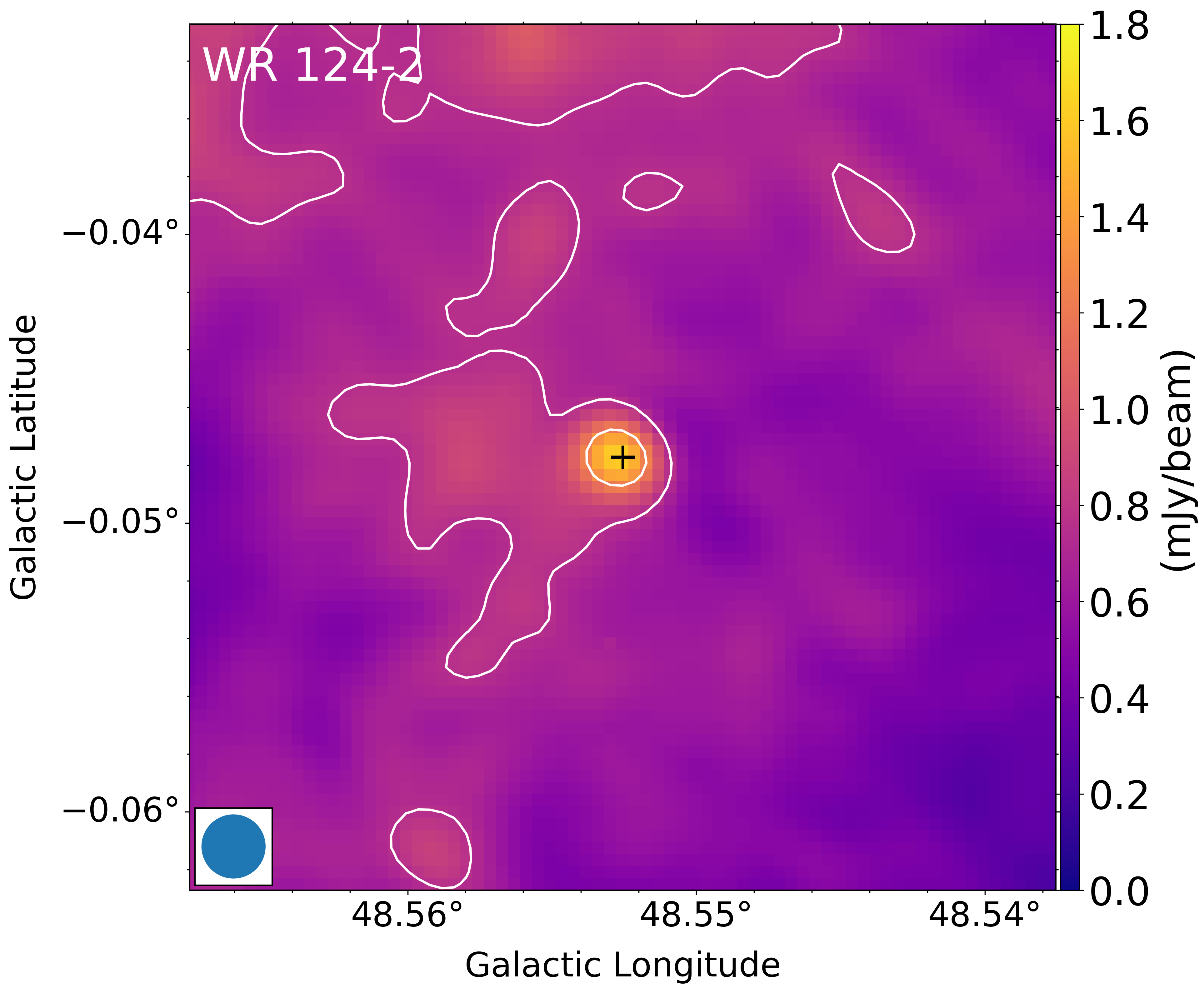} \\

\includegraphics[width=0.32\textwidth]{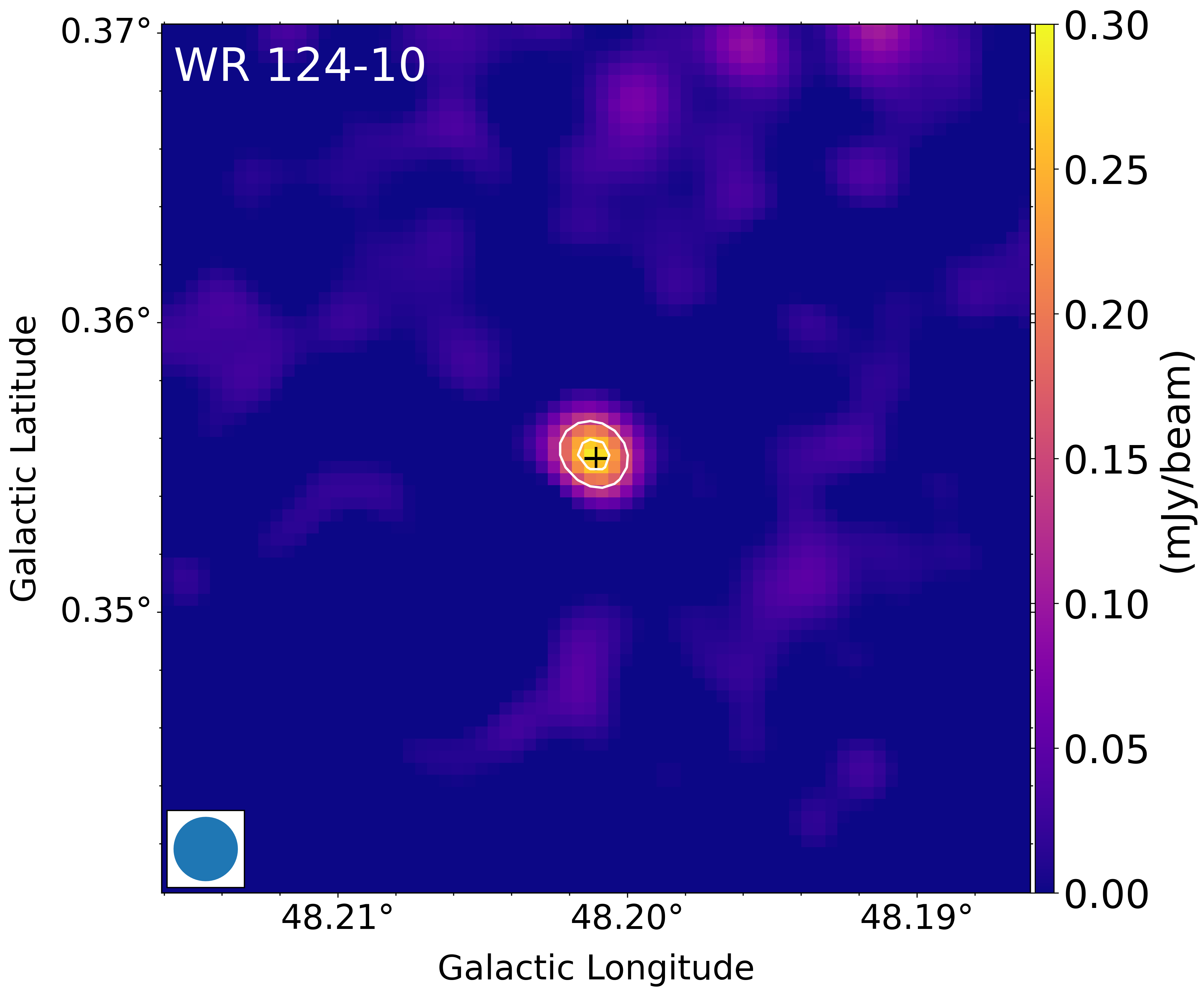} &
\includegraphics[width=0.32\textwidth]{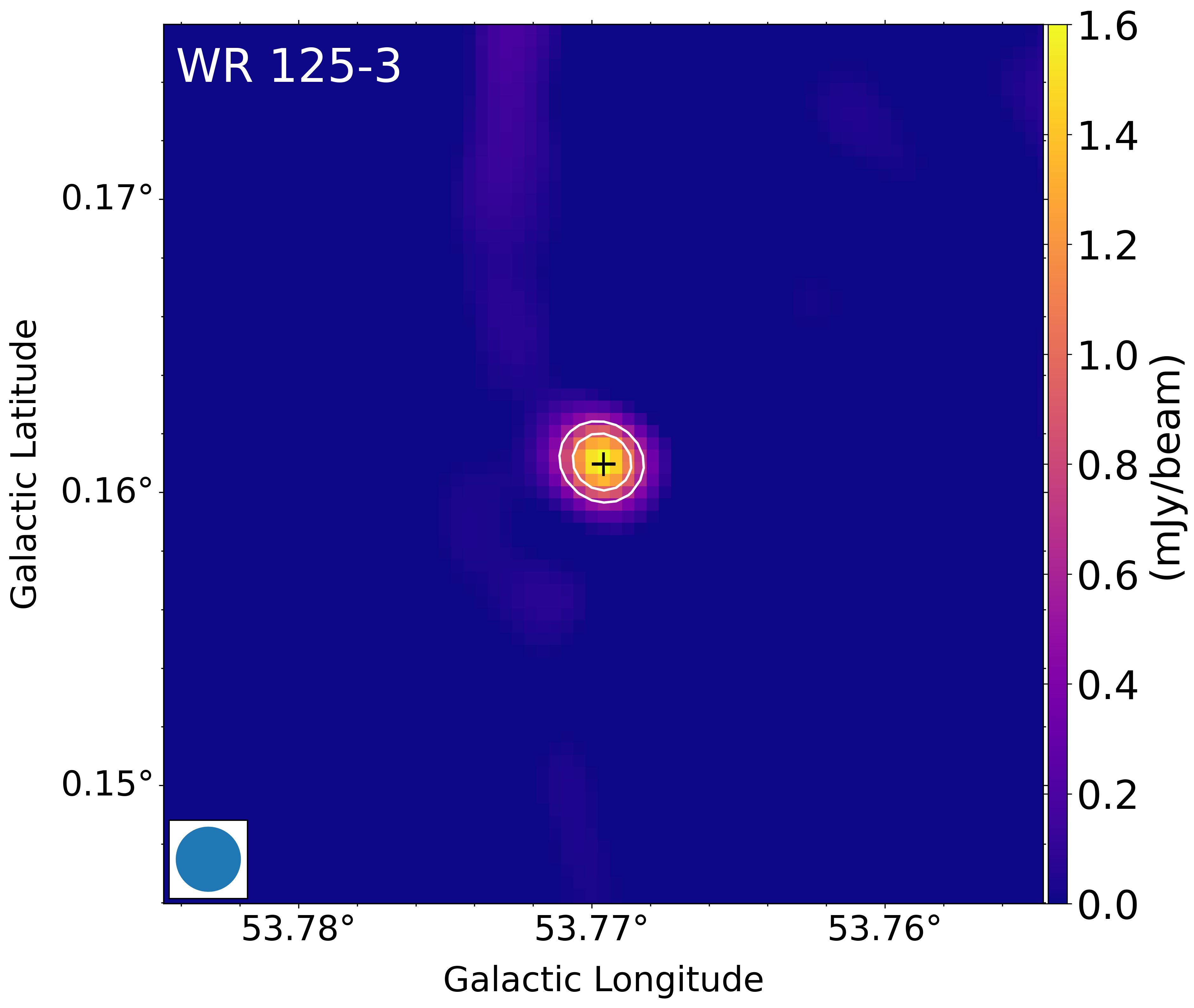} &
\\
\end{tabular}

\caption[]{Continued.}
\end{figure*}
\end{appendix}

\end{document}